\documentclass[twocolumn,showpacs,showkeys,aps,prb,superscriptaddress,amsmath,amssymb]{revtex4}
\usepackage{graphicx}
\usepackage{color}
\usepackage{amssymb,amsmath}
\usepackage[squaren]{SIunits}

\begin{document}

\title{Two-resonator circuit QED: Dissipative Theory}

\author{Georg M. Reuther}
\email{georg.reuther@physik.uni-augsburg.de}
\affiliation{Institut f\"ur Physik, Universit\"at Augsburg,
       Universit\"atsstra{\ss}e~1, D-86135 Augsburg, Germany}
\author{David Zueco}
\email{David.Zueco@physik.uni-augsburg.de}
\affiliation{Institut f\"ur Physik, Universit\"at Augsburg,
       Universit\"atsstra{\ss}e~1, D-86135 Augsburg, Germany}

\author{Frank Deppe}
\affiliation{Walther-Mei{\ss}ner-Institut, Bayerische Akademie der
  Wissenschaften, Walther-Mei{\ss}ner-Str. 8, D-85748 Garching, Germany}

\author {Elisabeth Hoffmann}
\affiliation{Walther-Mei{\ss}ner-Institut, Bayerische Akademie der
  Wissenschaften, Walther-Mei{\ss}ner-Str. 8, D-85748 Garching, Germany}

\author{Edwin P. Menzel}
\affiliation{Walther-Mei{\ss}ner-Institut, Bayerische Akademie der
  Wissenschaften, Walther-Mei{\ss}ner-Str. 8, D-85748 Garching, Germany}

\author{Thomas Wei{\ss}l}
\affiliation{Physik-Department, Technische Universit\"at M\"unchen,
  James-Franck-Str., 85748 Garching, Germany}

\author{Matteo Mariantoni}
\thanks{Corresponding address: Department of Physics, University of California, Santa Barbara, California 93106,
USA}
\affiliation{Walther-Mei{\ss}ner-Institut, Bayerische Akademie der
  Wissenschaften, Walther-Mei{\ss}ner-Str. 8, D-85748 Garching, Germany}
\affiliation{Physik-Department, Technische Universit\"at M\"unchen,
  James-Franck-Str., 85748 Garching, Germany}

\author{Sigmund Kohler}
\affiliation{Instituto de Ciencia de Materiales de Madrid, CSIC,
   Cantoblanco, E-29049 Madrid, Spain}

\author{Achim Marx}
\affiliation{Walther-Mei{\ss}ner-Institut, Bayerische Akademie der
  Wissenschaften, Walther-Mei{\ss}ner-Str. 8, D-85748 Garching, Germany}

\author{Enrique Solano}
\affiliation{Departamento de Qu\'{\i}mica F\'{\i}sica, Universidad del Pa\'{\i}s Vasco -
Euskal Herriko Unibertsitatea, Apdo. 644, 48080 Bilbao, Spain}
\affiliation{IKERBASQUE, Basque Foundation for Science, Alameda Urquijo 36, 48011 Bilbao,
Spain}

\author{Rudolf Gross}
\affiliation{Walther-Mei{\ss}ner-Institut, Bayerische Akademie der
  Wissenschaften, Walther-Mei{\ss}ner-Str. 8, D-85748 Garching, Germany}

\author{Peter H\"anggi}
\affiliation{Institut f\"ur Physik, Universit\"at Augsburg,
       Universit\"atsstra{\ss}e~1, D-86135 Augsburg, Germany}

\date{\today}

\begin{abstract}
We present a theoretical treatment for the dissipative two-resonator
circuit quantum electrodynamics setup referred to as quantum
switch. There, switchable coupling between two superconducting 
resonators is mediated by a superconducting qubit operating in the
dispersive regime, where the qubit transition frequency is far
detuned from those of the resonators. We derive an effective
Hamiltonian for the quantum switch beyond the rotating wave
approximation and provide a detailed study of the dissipative
dynamics. As a central finding,
we derive analytically how the qubit affects the quantum switch even
if the qubit has no dynamics, and we estimate the strength of this 
influence. The analytical results are corroborated by numerical
calculations, where coherent oscillations between the resonators, 
the decay of coherent and Fock states, and the decay of
resonator-resonator entanglement are studied. Finally, we suggest an
experimental protocol for extracting the damping constants of qubit
and resonators by measuring the quadratures of the resonator fields. 
\end{abstract}

\pacs{84.30.Bv, 03.67.Lx, 32.60.+i, 42.50.Pq}

\keywords{microwave switches, perturbation theory, quantum computing,
  quantum electrodynamics, superconducting cavity resonators,
  superconducting integrated circuits, superconducting switches} 

\maketitle

% *** SECTION ***
\section{Introduction\label{sec:intro}}
Circuit quantum
electrodynamics~\cite{Wallraff2004a,Blais2004a,You2005a} (QED) 
is the solid-state analog of quantum-optical cavity
QED.~\cite{Thompson1992a,Mabuchi2002a,Haroche2006a} While in the
latter natural atoms are coupled to 3D-cavities, the former is based  
on superconducting quantum circuits and the roles of the atoms and
the cavities are played by
qubit~\cite{Makhlin2001a,Wendin2006a,Clarke2008a} and microwave  
resonator circuits~\cite{Goeppl2008a,Niemczyk2009a,Lindstrom2007a},
respectively. In fundamental research, circuit QED architectures
have proved to be valuable for implementing quantum optics on a chip, for
which a rich toolbox has been
developed.~\cite{You2003a,Liu2006b,Astafiev2007a,Houck2007a,Leek2007a,%
Schuster2007a,Bishop2009b,Deppe2008a,Fink2008a,Hofheinz2009a}
These experiments were based on a single qubit coupled to a single
resonator. With applications for quantum information processing in
mind, an extension to multiple qubits seems natural.~\cite{Majer2007a,%
Sillanpaa2007a,DiCarlo2009a,Fink2009a} More recently, the potential
of using multiple resonators has been pointed out by several
authors.~\cite{Mariantoni2008a,PengXue2008a,Helmer2009a}

Under opportune circumstances, in a two-resonator circuit QED setup, a
superconducting qubit acts as a quantum switch between two
superconducting on-chip resonators.~\cite{Mariantoni2008a} To this
end, the qubit must be detuned from both resonators. The resulting
effective Hamiltonian describes a resonator-resonator interaction
whose coefficient has two contributions. The first contribution
depends on the qubit state and the qubit-resonator detuning and can
have a positive or negative sign. The second one has a definite sign
and stems from the fact that qubit and resonators are not point-like 
objects but extended circuits. Provided that the qubit always is in a
suitable energy eigenstate, the switch is turned off when both terms
are balanced and turned on otherwise. Beyond this simple protocol, the 
``quantumness'' of the setup can be exploited by bringing the qubit
into a superposition state with the resonators. This allows for
generating bi- and tripartite entanglement or Schr\"odinger cat
states.

In a real experiment, one expects the operation of the two-resonator
circuit QED setup to be affected by the various decoherence rates of
qubit and resonators. Since most implementations of superconducting
qubits can be tuned by external parameters, those rates depend not
only on the qubit type, but also on the operating point. So
far,~\cite{Bertet2005a,Yoshihara2006a,Deppe2007a,Kakuyanagi2007a,Houck2008a} 
they have been in the range of approximately
$1$-$200\,\mega\hertz$. In the case of the quantum switch,
only qualitative estimates on the effect  of qubit dephasing
exist.~\cite{Mariantoni2008a} However, a detailed quantitative
understanding of the possible effects stemming from the various
existing decoherence channels is indispensable for successful
experimental implementation. In particular, it is essential to analyze 
the effect of qubit decoherence sources on the coupled resonator
pair. Hence, in this work, we develop a complete dissipative theory 
for a circuit QED setup consisting of two resonators both dispersively
coupled to a single qubit. As a central result, we demonstrate that
qubit relaxation affects the resonators in second dispersive order,
whereas dephasing becomes an issue only in fourth dispersive order. 

The paper is structured as follows. In Sec.~\ref{sec:II}, we introduce
the system Hamiltonian and add the baths causing dissipation to the
system. We model the bath influence using a Bloch-Redfield quantum
master equation. Next, in Sec.~\ref{sec:beyondRWA}, we derive an effective
system Hamiltonian beyond the rotating-wave approximation (RWA). We
show that this extension results in quantitative, but not qualitative
changes compared to a treatment within RWA. Furthermore,  in
Sec.~\ref{sec:effqme}, we derive a simplified effective quantum master
equation suitable for analytical treatments. In particular, we use the
latter result to compute an explicit expression for the influence of  
the qubit dissipation channel on the two-resonator
system. Sec.~\ref{sec:results} contains numerical results for various    
prototypical operation modes of the quantum switch setup. These
include coherent oscillations between the resonators, the decay of
Fock and coherent states, and the decay of resonator-resonator
entanglement. We show that the agreement with the analytical results
obtained by means of the effective quantum master equation of
Sec.~\ref{sec:effqme} is excellent. Most importantly, we show that
qubit dissipation affects the switch only in second dispersive
order. Finally, in Sec.~\ref{sec:corr}, we suggest a protocol to
extract the damping constants of the system by measuring the field
modes of the resonators. The appendices contain technical details
about the calculations presented in this article.

% *** SECTION ***
\section{Dissipative two-resonator circuit QED\label{sec:II}}

\begin{figure}
  \centering
  \includegraphics[width=.99\columnwidth]{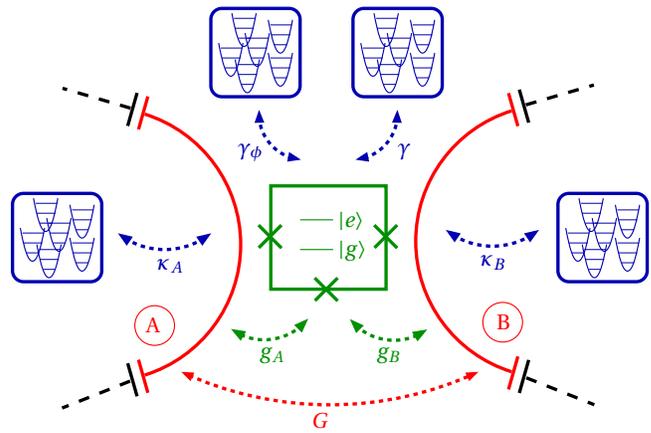}
  \caption{(Color online) Sketch of the two-resonator circuit QED
    system under analysis, including schematically the
    interaction with bosonic heat baths (blue boxes with oscillator
    potentials). Both microstrip or coplanar waveguides could be
    employed as resonators (red lines). As in the text, the 
    coupling qubit is exemplarily depicted as persistent-current flux
    qubit (green loop). The system is coupled to external circuits via
    coupling capacitors. The decay rates $\kappa_{\{A,B\}}$, $\gamma$
    and $\gamma_\phi$ are defined in Sec.~\ref{sec:qubitdecayrates}.}   
  \label{fig:sketch}
\end{figure}

We introduce a dissipative description for a circuit
QED architecture consisting of two on-chip microwave resonators that
are simultaneously coupled to one superconducting qubit. This setup is
sketched in Fig.~\ref{fig:sketch}. We emphasize that our formalism is
general in the sense that qubit and resonators can be based on any
suitable quantum circuits. However, whenever we need to give numbers,
we assume a persistent-current flux
qubit~\cite{Mooij1999a,Orlando1999a} coupled to two transmission line
resonators henceforth.  

% *** SUBSECTION ***
\subsection{System-bath model\label{sec:systembath}}

First, we write down the two-resonator circuit QED
Hamiltonian. A detailed derivation is given in
Ref.~\onlinecite{Mariantoni2008a}. The natural reference frame is the
laboratory basis, the physical basis of circuits and fields. 
\begin{equation}
\label{eq:HSphys}
\begin{split}
  \mathcal {H'}  = &\frac{\hbar \varepsilon}{2}
  \sigma '_z + \frac{\hbar \delta_Q}{2} \sigma '_x +
  \hbar \Omega _A a^\dagger a
  + \hbar \Omega _B b^\dagger b \\
  & + \hbar G (a + a^\dagger)(b + b^\dagger) \\
  & + \hbar g_A \sigma '_z (a + a^\dagger) + \hbar g_B
  \sigma '_z (b + b^\dagger) \, .
\end{split}
\end{equation}
The first two terms in the first line of the above
Hamiltonian represent the qubit in terms of the standard Pauli
operators $\sigma ' _x$ and $\sigma'_z$. The controllable energy bias
is $\hbar\varepsilon$, and $\hbar\delta_Q$ denotes the minimum level
splitting. In the particular case of a flux
qubit,~\cite{Mooij1999a,Orlando1999a} 
$\hbar\delta_Q$ is the tunnel splitting, and the energy bias
$\hbar\varepsilon = 2i_Q(\Phi_x^{\rm DC}{-}\Phi_0/2)$ can be tuned
by an externally applied flux $\Phi_x^{\rm DC}$. The quantities
$i_Q$ and $\Phi_0 = h/2e$ denote the qubit persistent current and the
magnetic flux quantum, respectively. When $\Phi_x^{\rm DC} = \Phi_0/2$
or, equivalently, $\varepsilon = 0$, the qubit is said to be biased at
its degeneracy or optimal point, where it is protected from
low-frequency noise to first order. The last two terms in the first
line of Eq.~(\ref{eq:HSphys}) represent the two resonators with
frequencies $\Omega_A$ and $\Omega_B$. Here, $a, b$ and
$a^{\dagger},b^{\dagger}$ are the annihilation and creation 
operators of the modes in resonators A and B, respectively. The second
line of Eq.~(\ref{eq:HSphys}) describes the geometric coupling between
the resonators, which is due to the fact that we are dealing with
circuits. The coupling coefficient $G$ contains contributions both
from a direct coupling and an interaction that is mediated by the
qubit circuit. Finally, the third line of
Eq.~(\ref{eq:HSphys}) describes the qubit-resonator coupling terms
with coefficients $g_A$ and $g_B$. As explained in
Sec.~\ref{sec:beyondRWA}, they give rise to a ``dynamical''
resonator-resonator coupling under appropriate conditions.

In a real experimental scenario, the two-resonator circuit is
unavoidably coupled to an external circuit that is characterized by an
impedance $Z(\omega)$. In a quantum mechanical description,
this impedance can be modeled by coupling the circuit bilinearly to
the modes of an electromagnetic environment consisting of
an infinite set of harmonic oscillators.~\cite{Yurke1984a,Ingold1992a}
Following 
this route, we obtain a Caldeira-Leggett-type system-bath
Hamiltonian,~\cite{Caldeira1983a,Hanggi1990a, Weiss1993a}
\begin{align}
  \label{eq:Hsb}
\nonumber
\mathcal H ' _\mathrm{tot}
=& \mathcal  H ' + 
\sum_\mu  Q_\mu \sum_j c_j ^\mu \left(d_{j,\mu}^\dagger + 
d_{j,\mu} ^{\vphantom{\dagger}} \right) \\
&+ \sum_\mu \sum_j \hbar \omega_j ^\mu
\left(d_{j,\mu}^\dagger  d_{j,\mu} ^{\vphantom{\dagger}} +
  \frac{1}{2} \right) 
\end{align}
The indices $\mu  \in \{A,B,x,z\}$ label the system-bath coupling
operators with respect to the different reservoirs the system
is coupled to. In detail,
\begin{align}
  \label{eq:coupling}
  \nonumber Q_A & =  (a+a^\dagger)\, , \qquad
   Q_x  = \sigma '_x \\
   Q_B & =  (b+b^\dagger)\, , \qquad
  Q_z  =  \sigma '_z \, .
\end{align}
The coupling coefficients $c^\mu_j$ represent the interaction between
the system and the different bath modes with frequencies
$\omega^\mu_j$, which are described by the bosonic annihilation and
creation operators $d_{j,\mu}^{\vphantom{\dagger}},d_{j,\mu}^{\dagger}
$. Within the scope of this paper, we consider the noise sources to be 
uncorrelated. This is justified since the different types of noise are
caused by fluctuations of distinct nature. In other words, we assume
that the baths are independent, $[d_{i,\mu}^{\vphantom{\dagger}},
d_{j,\nu}^\dagger] = \delta_{ij} \delta_{\mu \nu}$. We find it
noteworthy to mention that for $\mu \in \{x,z\}$ the coefficients
$c_j^\mu$ depend on the specific implementation of the qubit . For a 
flux qubit, the dominant noise source is believed to be flux
noise,~\cite{Yoshihara2006a,Deppe2007a,Kakuyanagi2007a} which
couples to the circuit via the $z$-axis in the laboratory frame.

In order to get more physical insight, we rotate $ \mathcal H '$ into
the qubit energy eigenbasis~\cite{Cohen-Tannoudji1992a} $\{ |g
\rangle, |e\rangle \}$, where $|g\rangle$ and $|e\rangle$ denote the
flux-dependent qubit ground and excited state,
respectively. Using the redefined Pauli operators 
\begin{equation}
  \begin{split}
    \sigma_x & = |g\rangle\langle e| + |e\rangle\langle g| =
    \cos\theta\,\sigma'_x - \sin\theta\,\sigma'_z
    \\
    \sigma_z & = |e\rangle\langle e| - |g\rangle\langle g| =
    \sin\theta\,\sigma'_x + \cos\theta\,\sigma'_z \; ,
  \end{split}
\label{eq:pauli}
\end{equation}
we obtain
\begin{align}\label{eq:HS}
\nonumber \mathcal H  & = \frac{\hbar\omega
    _\mathrm{qb}}{2} \sigma_z + \hbar \Omega _A a^\dagger a
  + \hbar \Omega _B b^\dagger b \\
\nonumber & + \hbar G (a + a^\dagger)(b + b^\dagger) \\
\nonumber  & + \hbar g_A ( \cos\theta\,\sigma_z {-} \sin\theta\,\sigma_x)
  (a +  a^\dagger) \\
  &+ \hbar g_B ( \cos\theta\,\sigma_z {-} \sin\theta\,\sigma_x) (b +
  b^\dagger) \, .
\end{align}
The flux-dependence is now encoded in the qubit energy level splitting
$\hbar\omega_\text{qb} = \hbar (\delta_Q^2+\varepsilon^2)^{1/2}$ and
the mixing angle $\theta = \arctan(\delta_Q/\varepsilon)$. The
qubit-bath coupling operators are rewritten as 
\begin{equation}
  \label{eq:coupling2}
    \begin{split}
      Q_x &= \sigma'_x = \cos\theta\,\sigma_x + \sin\theta\,\sigma_z
      \\
      Q_z &= \sigma'_z = \cos\theta\,\sigma_z - \sin\theta\,\sigma_x \,
      . 
    \end{split}
\end{equation}
They are defined along the rotated axes determined by the tunneling
matrix element $\hbar\delta_Q$ in $\sigma '_x$-direction,
and the energy bias $\hbar\varepsilon$ in $\sigma '_z$-direction. 
The system-bath interaction is fully characterized by the spectral
densities
\begin{equation}
  \label{JZ}
  J_\mu(\omega)=\sum_j|c_j^\mu|^2\delta(\omega_{j}-\omega)\, .
\end{equation}
In the case where decoherence is mainly caused by external
  circuitry, the spectral densities are proportional to the real part
  of the impedances $\mathrm{Re}[Z_\mu(\omega)]$. In general, internal
  loss mechanisms are also relevant in superconducting resonators at
  low powers and low temperatures. They often originate from
  fluctuators on the resonator surface, which are usually modeled as
  two-level systems. Thus, we interpret the $J_\mu (\omega)$ in an
  effective sense in that they include both the effects of external
  circuitry and internal losses. Our effective description does not
  cover the so-called excess phase noise though, i.\,e. low-frequency
  fluctuations in the resonator frequency itself, which originate from
  the surface fluctuators as well. As it was pointed out and
  investigated experimentally,~\cite{Gao2007a,Gao2008a} this leads to
  resonator dephasing. While such effects are not included in our
  modeling of decoherence, we cannot ensure that they will only be of
  minor importance with respect to operating  the two-resonator setup
  (see below in Sec.~\ref{sec:quantumswitch}).   
  In most experimental situations, however, decoherence is
  predominantly governed by external resonator losses. The
  corresponding external quality factor is characterized by the
  coupling capacitors to external circuitry. We note that resonator
  dephasing was not reported to play a major role in recent circuit
  QED experiments done with comparable resonators. In any case, the
  role of non-vanishing excess phase noise requires a separate, more
  detailed treatment with respect to an intended experimental
  realization of our setup.

% *** SUBSECTION ***
%
\subsection{Bloch-Redfield quantum master equation}
\label{sec:blochredfield}

The dissipative dynamics of the qubit-two-resonator system is obtained
by tracing out the bath degrees of freedom of the total density
operator $\varrho_\mathrm{tot}$ associated with the transformed
system-bath Hamiltonian,

\begin{align}
  \label{eq:Hsbeigen}
\nonumber
\mathcal H _\mathrm{tot}
=&  \mathcal  H + 
\sum_\mu  Q_\mu \sum_j c_j ^\mu \left(d_{j,\mu}^\dagger + 
d_{j,\mu} ^{\vphantom{\dagger}} \right) \\
&+ \sum_\mu \sum_j \hbar \omega_j ^\mu
\left(d_{j,\mu}^\dagger  d_{j,\mu} ^{\vphantom{\dagger}} +
  \frac{1}{2} \right)
  \, , 
\end{align}
where the qubit-bath coupling operators $Q_x$ and $Q_z$ are now
written in the qubit eigenbasis according to Eq.~\eqref{eq:coupling2}. 
For weak system-bath interaction, the baths can be eliminated within
Bloch-Redfield theory \cite{Redfield1957a,Blum1996a} as follows:
Assuming that the baths are initially in thermal equilibrium at
temperatures $T_\mu$ and not correlated with the system state
$\varrho$, the total system-bath state can be written as
$\varrho_\mathrm{tot}{\propto}\varrho \otimes\prod_\mu\exp(-\sum_j
\hbar\omega_j^{\mu} d_{j,\mu}^\dagger
d_{j,\mu}^{\vphantom{\dagger}}/k_\mathrm{B} T_\mu)$. Then, one can derive
within perturbation theory the quantum master equation for the reduced
system density operator
$\varrho= \mathrm{Tr}_\mathrm{bath}[\varrho_\mathrm{tot}]$.
This procedure yields

\begin{align}
  \label{eq:blochredfield}
\dot \varrho (t) = 
&- \frac{i}{\hbar} \big[ \mathcal H,  \varrho (t)  \big] \\ 
  \nonumber & +  \frac{1}{\hbar^2} \sum_\mu 
  \int_0 ^ \infty \mathrm{d} \tau K_\mu (\tau) \\
  \nonumber  &\times   \Big[ \tilde Q_\mu (-\tau)  \varrho (t)
  Q_\mu    - Q_\mu \tilde Q_\mu (-\tau) \varrho (t) \Big] \,
  +\, \mathrm{h.c.}    
 \end{align}
The environment correlation functions $K_\mu (\tau)$ are given by 

\begin{equation}
  \label{eq:kernel}
  K_\mu(\tau) = \frac{\hbar}{\pi} \int_0^\infty\!\!\!\!\mathrm{d} \omega
  J_\mu (\omega) \left[\coth\!\left(\frac{\hbar 
  \omega}{2 k_\mathrm{B}T_\mu}\right)\!\cos \omega\tau{-}
i\sin\omega\tau\right] \, ,
\end{equation}
where, $J_\mu(\omega)$ are the spectral densities~\eqref{JZ}. The
Heisenberg operators $\tilde Q_\mu(\tau) = U_0^\dagger(\tau) Q_\mu
U_0(\tau)$ are constructed via the system propagator
$U_0(\tau) = \mathcal
T[\exp\{-(i/\hbar)\int_0^{\tau}\mathrm{d}t\mathcal{H}(t)\}]$.
Here, the time ordering operator $\mathcal T$ is only
required for an explicitly time-dependent system Hamiltonian.

We note that Eq.~\eqref{eq:blochredfield} is based on a Born-Markov
approximation, since the bath correlation functions are supposed to
decay sufficiently fast as compared to typical timescales of intrinsic
system evolution. Thus, it was appropriate to extend the integral in
Eq.~\eqref{eq:blochredfield} to infinity. Consistently, we assume 
Ohmic spectral densities in the correlation functions of
Eq.~\eqref{eq:kernel}, modeling $Z(\omega)$ as an effective
resistance. However, this restriction is only necessary 
in the low-frequency region of the qubit environments. There, we
assume  
 \begin{equation}
   \label{eq:ohmic}
  J _\mu (\omega)=\alpha_\mu\omega
  \qquad\mu \in \{x,z\},\;\omega{\ll}\omega_\mathrm{qb}
 \end{equation}
and the coefficients $\alpha_\mu$ represent the 
dimensionless damping strengths. As we will see later, in the
high-frequency regime, we are interested only in infinitely small
intervals around frequencies such as $\omega_\mathrm{qb}$,
$\Omega_A$, and $\Omega_B$. Hence, the Born-Markov approximation
remains justified by expanding $J_{\mu}(\omega)$
to first order in these intervals. In this way, the only remaining
restriction is that $J_\mu (\omega)$ is a smooth function 
around the frequencies of relevance. Within the scope of this work, we
shall consider Eqs.~\eqref{eq:blochredfield}--\eqref{eq:ohmic} as a
full description of the influence of dissipation and decoherence on
the two-resonator setup. 

This reasoning excludes in particular
$1/f$-noise, which affects the phase coherence of superconducting 
qubits~\cite{Ithier2005a,Yoshihara2006a,Kakuyanagi2007a} due to its
high impact at low frequencies. One typically describes $1/f$-noise by 
calculating the accumulated random phase as a function of time for
specific experimental
protocols.~\cite{Ithier2005a,Deppe2007a,Schriefl2006a} However, as 
shown below, we expect the effect of qubit dephasing to be
suppressed even more than relaxation effects in the setup described
here.

Since the quantum master equation~\eqref{eq:blochredfield} is 
non-trivial with respect to analytical treatment, we only use it for
numerical purposes. However, in Sec.~\ref{sec:dissipation-disp}, we 
derive a simplified effective quantum master equation in the
dispersive regime, which will allow for analytic insight into the
dissipative behavior of the two-resonator circuit.

% *** SUBSECTION ***
\subsection{Qubit decay rates}
\label{sec:qubitdecayrates}

So far, we have modeled the coupling of the qubit to the baths in the
laboratory frame. In this way, we can include the relevant
noise channels for any particular qubit architecture into our
formalism easily. However, with regard to physical understanding, it
is more favorable to work in the qubit energy eigenframe and refer to 
what is commonly called energy relaxation and pure dephasing. The
former describes bath-induced level transitions, while the latter
accounts for the pure 
loss of phase coherence without a change of the system energy. In
order to define the decay rates corresponding to these two
processes, we first review the dissipation mechanisms of the qubit
alone. To this end, we derive the quantum master equation describing
the evolution of the reduced qubit density matrix
$\varrho_\mathrm{qb}$ for a single qubit associated with the
Hamiltonian
$\mathcal{H}_\mathrm{qb} = \hbar\omega_\mathrm{qb}\sigma_z/2$ in the
energy eigenbasis. Using the formalism detailed in
App.~\ref{sec:qubitqme}, we find    
\begin{equation}\label{eq:qubitQME}
  \begin{split}
    \dot \varrho_\mathrm{qb}(t) =
    & {}-\frac{i}{\hbar} 
    \big[\mathcal H _\mathrm{qb},\varrho_\mathrm{qb} (t) \big]\\
    & {}+ \gamma(\omega_\mathrm{qb})\left(\sigma^- \varrho_\mathrm{qb} (t)
      \sigma^+ - \frac{1}{2} \big[\sigma^+ \sigma^-,\varrho_\mathrm{qb}
      (t)  \big]_+\right)  \\
    & {}+ \gamma_\phi(\omega{\rightarrow}0) \big(\sigma_z
      \varrho_\mathrm{qb} (t) \sigma_z - \varrho_\mathrm{qb} (t)\big) \, ,
  \end{split}
\end{equation}
where $[\mathcal{A},\mathcal{B}]_+ = \mathcal{AB}{+}\mathcal{BA}$
denotes the anti-commutator between the operators $\mathcal{A}$ and
$\mathcal{B}$. The dissipator in the third line of
Eq.~\eqref{eq:qubitQME} does not affect the populations of the qubit
eigenstates, but only accounts for the decay of the
off-diagonal elements of the density operator. Thus, the rate
$\gamma_\phi(\omega{\rightarrow}0)$ can be associated with pure
dephasing. The dissipator in the second line of
Eq.~\eqref{eq:qubitQME} induces transitions between the qubit
eigenstates, hence $\gamma(\omega_\mathrm{qb})$
characterizes relaxation. Assuming an overall temperature
$T= T_x= T_z$, and following Eq.~\eqref{eq:relaxationrate1} 
and Eq.~\eqref{eq:dephasingrate1}, the 
qubit energy relaxation rate $\gamma(\omega_\mathrm{qb})$ and pure
dephasing rate $\gamma_\phi(\omega{\rightarrow}0)$ are obtained as 
\begin{align}
   \gamma(\omega_\mathrm{qb}) 
   &= J_x(\omega_\mathrm{qb})\cos^2\theta
   +J_z(\omega_\mathrm{qb})\sin^2\theta
   \label{eq:relaxationrate2} \\
   \gamma_{\phi}(\omega{\rightarrow}0)  
   &= \frac{k_\mathrm{B}T}{\hbar}
   (\alpha_z\cos^2\theta+\alpha_x\sin^2\theta)\,.
   \label{eq:dephasingrate2}
\end{align}
Equations~\eqref{eq:relaxationrate2} and
\eqref{eq:dephasingrate2} link the physical system-bath
interactions quantified in the laboratory frame to the pure
bit-flip and dephasing mechanisms relevant in the qubit
eigenbasis. Moreover, they highlight the dependence of the pure decay
rates on the applied flux in terms of the mixing angle
$\theta$. In particular, for a flux qubit,
flux noise can be responsible for both relaxation and
dephasing. We emphasize that, in this special scenario, 
$J_x(\omega)= 0$ and $J_z(\omega)\!\ne\!0$, and 
Eq.~\eqref{eq:relaxationrate2} is consistent with results from other 
works.~\cite{Ithier2005a,Schriefl2006a,Deppe2007a}

% *** SECTION ***
\section{Analytical treatment of decoherence in the dispersive limit}
\label{sec:dissipation-disp}

In the setup of Fig.~\ref{fig:sketch}, the
qubit can mediate a controllable coupling between the two resonators, 
i.\,e., it can act as a quantum switch between them. In this section,
we review the quantum switch Hamiltonian of
Ref.~\onlinecite{Mariantoni2008a} and extend it beyond the
rotating-wave approximation. Furthermore, we derive an effective
quantum master equation which allows us to understand by purely
analytical arguments that the quantum switch is affected by the qubit
dissipation only in second (relaxation) and fourth order (dephasing),
respectively.  
 
% *** SUBSECTION ***
\subsection{Dispersive Hamiltonian within the rotating wave approximation: The quantum switch\label{sec:quantumswitch}}

In order to function as a quantum switch, the two-resonator circuit
must be operated in the dispersive limit, where the qubit-resonator
detuning $\Delta$ is large as compared to the qubit-oscillator
coupling,  
\begin{equation}
   g\ll\Delta\,,\quad\Delta=\omega_\mathrm{qb}-\Omega\,,
   \label{eq:dispersive}
\end{equation} 
and the parameter $\lambda_\Delta$ is necessarily small:
\begin{equation}
  \lambda_\Delta = \frac{g \sin \theta}{\Delta}
   \,,\qquad|\lambda_\Delta|\ll1
   \,. 
\label{eq:lambdaX-}
\end{equation}
Here and henceforth, we confine ourselves to symmetric setups with
$\Omega = \Omega_{\{A,B\}}$ and $g = g_{\{A,B\}}$. This is not
expected to be a serious restriction in practice,
though.~\cite{Mariantoni2008a}    

In the dispersive limit determined by Eq.~\eqref{eq:dispersive}, the
Hamiltonian of Eq.~\eqref{eq:HS} can be diagonalized
approximately. To this end, it 
is first simplified with a rotating wave approximation as
follows. Writing $\sigma_x = \sigma^+ + \sigma^-$ with the fermionic
raising and lowering operators $\sigma^+ = |e\rangle\langle g|$ and
$\sigma^- = |g\rangle\langle e|$, one can move to
the interaction picture with respect to the uncoupled
Hamiltonian. Then, the coupling operators $\sigma^+ a$, $\sigma^-
a^\dagger$, $\sigma^+ b$, and $\sigma^- b^\dagger$ oscillate with the
phase factors $\exp [\pm i\Delta t]$, whereas $\sigma^- a$, $\sigma^+
a^\dagger$, $\sigma^- b$, and $\sigma^+ b^\dagger$ oscillate with
$\exp [\pm i\Sigma t]$, where 
\begin{equation}
   \Sigma=\Omega+\omega _\mathrm{qb} \, .
   \label{eq:sumfrequency}
\end{equation}
Close to resonance, the resonator-qubit detuning is small and,
consequently, $|\Delta| \ll \Sigma$. Thus, the former set of
operators oscillate slowly, whereas the latter exhibit fast
``counter-rotating'' oscillations. For sufficiently weak coupling
$g  \ll \min\{\omega_\mathrm{qb},\Omega\}$, one can
separate timescales and average the counter-rotating
terms to zero. In this way, the first-order interaction
Hamiltonian between qubit and resonators is
Jaynes-Cummings-like~\cite{Jaynes1963} and we describe our system
with 
\begin{equation}
  \begin{split}
  \mathcal{H}^\mathrm{RWA}={} &
  \frac{\hbar\omega_\mathrm{qb}}{2}\sigma_z
  +\hbar\Omega a^\dagger a+\hbar\Omega b^\dagger b
  \\
  & {}-\hbar\Delta\lambda_\Delta
  \big(\sigma^+a+\sigma^-a^\dagger+\sigma^+b+\sigma^-b^\dagger\big)
  \\
  & {}+\hbar G\big(a^\dagger b + ab^\dagger\big)\,.
  \end{split}
  \label{eq:HSRWA}
\end{equation}
In a second step, we apply the transformation
$\mathcal{U}^\mathrm{RWA}= \exp(-\lambda_\Delta\mathcal{D})$, where 
\begin{equation}
  \mathcal{D} = \sigma^-a^\dagger-\sigma^+a+\sigma^-b^\dagger-\sigma^+b\,.
  \label{eq:X-}  
\end{equation}
Finally, one truncates the transformed Hamiltonian
$\mathcal{H}_\mathrm{eff}^\mathrm{RWA} = 
\mathcal{U}^\dagger_\mathrm{RWA} \mathcal{H^\mathrm{RWA}} 
\mathcal{U}_\mathrm{RWA}$ to second order in $\lambda_\Delta$, 
yielding
\begin{equation}
  \label{HdiagRWA}
  \begin{split}
    \mathcal{H}_\mathrm{eff}^\mathrm{RWA} = \,
    & \hbar\Omega\big(a^\dagger a+b^\dagger b + 1\big )
    + \frac{\hbar \omega_\mathrm{qb}}{2} \sigma_z \\
    & {}+\hbar\Delta\lambda_\Delta^2\sigma_z
     \big(a^\dagger a + b^\dagger b+1\big)
    \\
    & {}+\hbar\hat{g}_\mathrm{SW}^\mathrm{RWA}
     \big(ab^\dagger + a^\dagger b\big) \,.
  \end{split}
\end{equation}
Here, the first line describes qubit and resonators, the second
ac-Stark/Zeeman and Lamb-shifts, and the third an effective coupling
between the two resonators. The corresponding coupling operator is  
\begin{equation}
  \label{eq:gsw}
  \hat{g}_\mathrm{SW}^\mathrm{RWA} = G+\Delta\lambda_\Delta^2\sigma_z\,.
\end{equation}
A remarkable feature of the Hamiltonian of Eq.~\eqref{HdiagRWA} is
that it commutes with $\sigma_{z}$, i.e.,
$[\mathcal{H}_\mathrm{eff}^\mathrm{RWA},\sigma_{z}]=0$. Consequently,
the qubit state will not change during the unitary evolution of the
system. When the qubit is prepared in a suitable eigenstate, it 
can be traced out. Throughout this work, we consider the
qubit to be initially prepared in its ground state ${|}g 
\rangle \langle g {|}$. Then, $\hat{g}_\mathrm{SW}$ simplifies
to the resonator-resonator coupling constant 
\begin{equation}
  \label{eq:switchsettingcondition}
  g_\mathrm{SW}^\mathrm{RWA} = G-\lambda_\Delta^2 \Delta\,.
\end{equation}
In this case, the Hamiltonian of Eq.~\eqref{HdiagRWA} describes two
coupled harmonic oscillators. By means of either adiabatic or
oscillating external flux signals, the qubit can be tuned such that
the interaction between the resonators is either switched on
($|g_\mathrm{SW}^\mathrm{RWA}| \neq 0$) or off
($|g_\mathrm{SW}^\mathrm{RWA}| = 0$). This feature is referred to as
the switch-setting condition. With the help of specific
protocols, it can be utilized to create entangled states out of
initial bi-resonator product states. 

We note that the effective coupling between both resonators can be
interpreted as a beam-splitter interaction. A comparable quantum
optical setup was proposed in Ref.~\onlinecite{Schneider1997a}. There,
an atom passing through a cavity serves to create entanglement between
two optical fields inside the  cavity. That system is described by an
effective Hamiltonian quite analogous to Eq.~\eqref{HdiagRWA}.

For this work the
``adiabatic'' shift protocol is of particular relevance. There,
parameters are initially chosen so as to fulfill the switch-setting
condition when the qubit is in ${|}g\rangle$. Then, the
resonator-resonator interaction can be turned on by adiabatically
varying the flux bias. Experiments have shown that a flux change slow
enough to avoid significant population of the excited state can be
realized easily even in pulsed setups.~\cite{Deppe2007a} 

Regarding the influence of a dissipative environment on the quantum
switch, we already note that, for suitable switching protocols and at
sufficiently low temperatures, the qubit energy relaxation and
dephasing will not affect the operation of the switch in first
order. As one of the main results of this work, we give analytic
arguments to put this statement on firm theoretical footings in 
Sec.~\ref{sec:effqme}.  

% *** SUBSECTION *** 
\subsection{Dispersive Hamiltonian beyond the rotating-wave approximation\label{sec:beyondRWA}} 

In the process of deriving Eq.~\eqref{HdiagRWA} in the previous
section, a rotating wave
approximation is applied to the Hamiltonian~\eqref{eq:HS}
at the level of first-order in the qubit-oscillator coupling.
However, it has recently been revealed that neglecting the
counter-rotating terms may lead to inaccuracies.~\cite{Zueco2009b}
Especially in the case of far detuning described by
Eq.~\eqref{eq:dispersive}, the rotating-wave approximation causes
noticeable deviations from results obtained numerically from the full
Hamiltonian of Eq.~\eqref{eq:HS} for typical parameters. Nevertheless,
the effective, dispersive Hamiltonian can be obtained
 by means of the unitary transformation     
\begin{equation}
\label{eq:Schrieffer1}
  \mathcal{U} = \exp\big({}-\lambda_\Delta \mathcal{D}
                -\lambda_\Sigma \mathcal{S}
                -\lambda_\Omega\mathcal{W}\big) \,.
\end{equation}
Here,
\begin{align}
\mathcal{S} &= \sigma^- a - \sigma^+ a^\dagger +\sigma^- b - \sigma^+ b
^\dagger  \label{eq:Y-} \\
\mathcal{W} &= \sigma_z (a - a^\dagger) + \sigma_z (b - b^\dagger) \, ,
 \label{eq:Z-}
\end{align}
and the corresponding coefficients are
\begin{align}
\lambda_\Sigma &=\frac{g \sin \theta}{\Sigma}
 \,,\qquad|\lambda_\Sigma|\ll1 
\label{eq:lambdaY-} \\
\lambda_\Omega &= \frac{g \cos \theta}{\Omega}
 \,,\qquad|\lambda_\Omega|\ll1\,.
 \label{eq:lambdaZ-}
\end{align}
The above inequalities allow us to discard terms of orders higher than 
$\lambda_{\{\Delta, \Sigma, \Omega\}}^2$ when 
computing the effective second-order Hamiltonian 
$\mathcal{H}_{\rm eff} = \mathcal{U}^\dagger\mathcal{H}\mathcal{U}$. 
In this case,  
\begin{equation}
  \label{Hdiag1}
  \begin{split}
    \mathcal H _{\rm eff} ={} 
    & \hbar\Omega\big(a^\dagger a+b^\dagger b+1\big)
    + \frac{\hbar\omega_\mathrm{qb}}{2}\sigma_z
    \\
    & {}+\hbar\big(\Delta\lambda_\Delta^2+\Sigma\lambda_\Sigma^2\big)
    \sigma_z \big (a^\dagger a + b^\dagger b + 1\big)
    \\
    & {}+ \hbar \hat g_\mathrm{SW}\big(a b^\dagger + a ^\dagger b\big)
  \end{split}
\end{equation}
becomes diagonal. In the above equation, we use the
qubit-state-dependent resonator-resonator coupling operator 
\begin{equation}
  \label{eq:switchmod}
  \hat{g}_\mathrm{SW} = 
  G+\big(\lambda_\Delta^2\Delta+\lambda_\Sigma^2\Sigma\big) \sigma_z\,.
\end{equation}

At this point, we can gain insight about the effect of the
transformation $\mathcal{U}$ [Eq.~\eqref{eq:Schrieffer1}] on the
Hamiltonian. We first note that, when applying the rotating-wave
approximation to $\mathcal{H}$, only the exponent $\mathcal{D}$ is
required to produce a diagonal second-order Hamiltonian,
cf. Sec.~\ref{sec:quantumswitch}.
Beyond this simple scenario,~\cite{Zueco2009b} the exponent
$\mathcal{S}$ cancels the first-order counter-rotating terms of
$\mathcal{H}$. Furthermore, the polaron transformation represented by
the exponent $\mathcal{W}$ must be applied to eliminate off-diagonal
interaction terms such as $g\cos\theta\sigma_z(a+a^\dagger)$, which
cause qubit-state-dependent energy shifts of the oscillator
coordinates when the qubit is biased away from its degeneracy point.  

However, in
$\mathcal{H}_\mathrm{eff}$, terms of the order $\lambda_{\{\Delta,
 \Sigma, \Omega\}}^2$, such as, e.\,g., 
$\sigma^+(a^{(\dagger)})^2$ or $\sigma^+a^\dagger a$, need to be
canceled with a rotating-wave argument. We emphasize that this
rotating-wave approximation in second-order in 
$\mathcal{H}_\mathrm{eff}$ still allows for an accurate description of
our system in the dispersive regime, whereas a rotating-wave
approximation in the first-order Hamiltonian $\mathcal{H}$ does
not. Following the same reasoning, we may also neglect terms $\propto
G/\Delta, G/\Sigma,\ldots \ll 1$. 
\begin{figure}
  \centering
  \includegraphics[scale=.9]{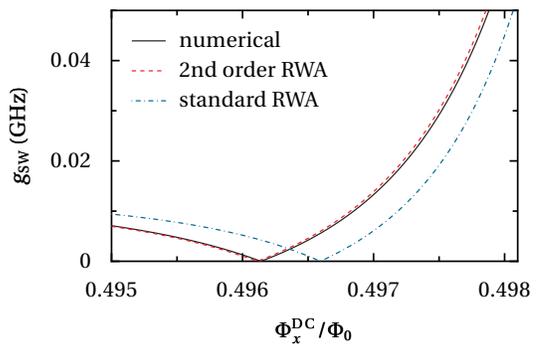}
  \caption{(Color online) Switch setting coefficient $g_\mathrm{SW}$
    (qubit in the ground state) for the RWA (blue dashed line) and
    non-RWA case, Eq.~\eqref{eq:switchmodgs} (red dashed line),
    compared to numerical data 
    obtained from the diagonalization of Hamiltonian (\ref{eq:HSphys})
    (black solid line). All quantities are plotted against the qubit flux
    bias $\Phi_x^{\rm DC}$. Parameters are $\Omega/2\pi=
    3.5\,\text{GHz}$, $\delta_Q/2\pi= 4\,\text{GHz}$, $G/2\pi=
    0.0022\,\text{GHz}$ and $g/2\pi= 0.24\,\text{GHz}$.}
  \label{fig:gsw}
\end{figure}
The effective Hamiltonian $\mathcal{H}_\mathrm{eff}$ of
Eq.~\eqref{Hdiag1} has the same structure as its rotating-wave
counterpart $\mathcal{H}_\mathrm{eff}^\mathrm{RWA}$ of
Eq.~\eqref{HdiagRWA}. However, there is one important quantitative
difference: the detuning dependence $\lambda_\Delta^2\Delta$ of the
coupling coefficients is replaced by the expression
$(\lambda_\Delta^2\Delta + \lambda_\Sigma^2\Sigma)$. In particular,
the effective resonator-resonator coupling constant is 
\begin{equation}
  \label{eq:switchmodgs}
  g_\mathrm{SW}=G-\big(\lambda_\Delta^2\Delta+\lambda_\Sigma^2\Sigma\big)\,
\end{equation}
for the qubit being in its ground state. The effect of the
counter-rotating terms is visualized in Fig.~\ref{fig:gsw}. There, we
compare $g_\mathrm{SW}^\mathrm{RWA}$ and $g_\mathrm{SW}$ to the
numerically exact coupling coefficient for adequate parameters. 
Obviously, in contrast to $g_\mathrm{SW}^\mathrm{RWA}$, the agreement
is excellent for $g_\mathrm{SW}$. This finding once more confirms the
necessity to include counter-rotating terms of first order in the 
qubit-oscillator coupling in the full system Hamiltonian. It also
confirms the validity of the rotating-wave approximation in second
order of $\lambda_{\{\Delta,\Sigma,\Omega\}}$ applied to the
dispersive Hamiltonian. We also illustrate the importance of the
non-RWA features below, where we develop a dissipative description of
the quantum switch  Hamiltonian coupled to different reservoirs.

% *** SUBSECTION ***
\subsection{Effective master equation for the quantum switch setup\label{sec:effqme}}

In this section, we analytically investigate the dissipative behavior
of the two-resonator-qubit system. To this end, we derive
an effective quantum master equation for the reduced
density matrix of our system in the dispersive
limit.~\cite{Boissonneault2009a,Boissonneault2009b} In particular,
we study the additional dissipation imposed on the resonators due to
the presence of the qubit. 

In principle, we combine the procedure explained in
Sec.~\ref{sec:beyondRWA} with that of Sec.~\ref{sec:qubitdecayrates}
and apply it to the system-bath Hamiltonian $\mathcal{H}_\mathrm{tot}$
of Eq.~\eqref{eq:HS}, which includes all counter-rotating
terms. First, we compute the total dispersive Hamiltonian
$\mathcal{H}_{\mathrm{tot,eff}}= \mathcal{U}^\dagger
\mathcal{H}_\mathrm{tot}\mathcal{U}$ using the
transformation~\eqref{eq:Schrieffer1} and truncate it to 
second order with respect to the parameters $\lambda_{\{\Delta, 
\Sigma,\Omega\}}$, as described above. During this
procedure, we obtain the effective system-bath coupling operators 
\begin{equation}
  \label{eq:transfcpl}
  Q _{\mu,\mathrm{eff}} = \mathcal U^\dagger Q_\mu \mathcal U \,
  , \quad \mu \in \left\{ A,B,x,z \right\} \, .
\end{equation}
The explicit expressions for these effective coupling operators are
given in App.~\ref{sec:effectivecoupling}. In the next step, we derive
the effective quantum master equation following the lines of
Refs.~\onlinecite{Breuer2001a,Scala2007a}. While
the interested reader can find the details in App.~\ref{sec:deriving},
we give a short summary of the most important steps in the
following. Motivated by the usual experimental conditions in circuit 
QED, we assume an equal temperature for all baths, $T =
T_{\{x,z,A,B\}}$, and confine ourselves to the low-temperature regime 
$k_\mathrm{B}T/\hbar \ll
\min\{\omega_\mathrm{qb},\Delta,\Omega,\Sigma\}$. Consequently, we
neglect all contributions to the dissipative system dynamics that
describe energy absorption from the baths. 

Using Eq.~\eqref{eq:blochredfield} as a starting point, we move first
to an interaction picture with respect to the uncoupled qubit and
resonators and insert the spectral decompositions of the effective
coupling operators. In the following, we perform a semi-secular
approximation.~\cite{Scala2007a} To this end, we dismiss 
terms that evolve rapidly compared to the time evolution of the system
state, i.\,e. on system timescales 
$\{\Omega,\omega_\mathrm{qb}, \Delta, \Sigma\}^{-1}$. On the
contrary, we keep those terms that oscillate slowly at
frequencies such as $\lambda_\Delta^2\Delta, \lambda_\Sigma^2\Sigma, 
\lambda_\Omega^2\Omega$. We emphasize that our result goes beyond the
standard Lindblad master equation, where one would perform a full
secular approximation, dismissing all oscillating contributions. 
In this way, we obtain the effective quantum
master equation for the reduced system state,
Eq.~\eqref{eq:effQME}. There, we assume
$\alpha_{\{x,z\}} \ll \hbar\omega_\mathrm{qb}/k_\mathrm{B}T$ in
order not to violate the Markov approximation. 

In order to gain physical insight into the influence of dissipation on
the quantum switch setup, we can simplify the complicated effective
master equation of Eq.~\eqref{eq:effQME}. In the dispersive regime,
the qubit mediates part of the coupling between the resonators by
exchanging virtual, but not real excitations with them. In particular,
as discussed in Sec.~\ref{sec:beyondRWA}, the switch can be operated
in a way that the qubit is initially prepared in the ground state and
remains there during the whole time evolution, as it cannot suffer
from further decoherence. In this scenario, the qubit degrees of
freedom can be traced out and the reduced Hamiltonian of the coupled
resonators becomes 
\begin{equation}
  \label{eq:Heffcav}
  \begin{split}
    \mathcal{H} _\mathrm{cav}^\mathrm{eff}={} 
    & \hbar\Omega\big(a^\dagger a+b^\dagger b+1\big)
    \\
    & {}-\hbar\left(\lambda_\Delta^2\Delta + 
      {\lambda_\Sigma^2\Sigma}\right)
      \big(a^\dagger a+b^\dagger b+1\big)
    \\
    & {}+\hbar g_\mathrm{SW}(a b^\dagger+a^\dagger b)\,. 
  \end{split}
\end{equation}
With the dissipator $\mathcal{D}[X]$ acting on an operator $X$ in the
product Hilbert space of the resonators, 
\begin{equation}
\label{eq:dissipator}
\mathcal{D}[X]\varrho_\mathrm{cav}=X\varrho_\mathrm{cav}X^\dagger
-\frac{1}{2}\big[X^\dagger X,\varrho_\mathrm{cav}\big]_+\,,
\end{equation}
we can write down the effective Lindblad-type quantum master equation
for the reduced state $\varrho_\mathrm{cav}$ of the two coupled
oscillators up to second order in $\lambda_\Delta$ and
$\lambda_\Sigma$, 
\begin{equation}
  \label{eq:effQMEcav}
  \begin{split}
    \dot\varrho_\mathrm{cav}={} & {}-\frac{i}{\hbar}
    \Big[\mathcal{H} _\mathrm{cav}^\mathrm{eff} ,
    \varrho_\mathrm{cav}^{\vphantom{\dagger}}\Big]
    \\
    & {}+\kappa_A\mathcal{D}\big[a\big]\varrho_\mathrm{cav}
    +\kappa_B\mathcal{D}\big[b\big]\varrho_\mathrm{cav}
    \\
    & {}+\kappa_\mathrm{qb}\mathcal{D}\big[a{+}b\big]\varrho_\mathrm{cav} \, .
  \end{split}
\end{equation}
The above equation reveals the relevant processes governing the
dissipative behavior of the quantum switch. The dissipators
$\mathcal{D}[a]$ and $\mathcal{D}[b]$ represent the independent decay
channels due to the individual environments of the resonators A and B,
respectively. The corresponding decay rates are the inverse lifetimes
of the uncoupled resonators, $\kappa_A= J_A(\Omega)$ and
$\kappa_B= J_B(\Omega)$. These rates may incorporate the
combined effects of internal and external loss mechanisms, according
to the discussion in Sec.~\ref{sec:systembath}.
In addition to these contributions, the
qubit introduces extra dissipation on the resonators via the
dissipator $\mathcal{D}[a + b]$. The appearance of the center-of-mass
coordinate $a{+}b$ of the two-resonator system in the dissipator
originates from the system Hamiltonian of Eq.~\eqref{eq:HS}, where
the qubit couples to the resonator ``center of mass'' coordinate,
i.\,e. the interaction is proportional to 
$\sigma_x(a + b + a^\dagger + b^\dagger)$ and
$\sigma_z(a + b + a^\dagger + b^\dagger)$, respectively. The
qubit-induced damping rate is 
\begin{equation}
  \begin{split}
    \kappa_\mathrm{qb}
    &{}=\left(\lambda_\Delta{+}\lambda_\Sigma\right)^2
    \left(J_x(\Omega)\cos^2\theta+J_z(\Omega)\sin^2\theta\right)
    \\
    &{}=\left(\lambda_\Delta{+}\lambda_\Sigma\right)^2\gamma(\Omega)\,,
  \end{split}
\label{eq:gammaeff}
\end{equation}
where $\gamma(\Omega)$ is the rate defined in
Eq.~\eqref{eq:relaxationrate1} for the bare qubit. In the expressions
for $\kappa_A$, $\kappa_B$, and $\kappa_\mathrm{qb}$, the spectral
densities $J_{\{A,B,x,z\}}(\omega)$ are required to be smooth
functions at $\omega= \Omega$ in order that Ohmic
behavior can be assumed locally. The reasoning is the same as the one
presented in App.~\ref{sec:qubitqme}. 

The qubit-induced damping rate of the two-resonator system,
$\kappa_\mathrm{qb}$ of Eq.~\eqref{eq:gammaeff}, constitutes one
central result of this work and has several remarkable features. First
of all, we note that $\gamma(\Omega)$ has the same functional
dependence on the qubit mixing angle $\theta$ as the relaxation rate 
$\gamma(\omega_\mathrm{qb})$ of the bare qubit,
Eq.~\eqref{eq:relaxationrate2}. However, $J_{\{A,B,x,z\}}(\Omega)$ and 
the corresponding $J_{\{A,B,x,z\}}(\omega_\mathrm{qb})$ are not
necessarily equal, thus the values of both rates are different in
general. Second, the rate $\kappa_\mathrm{qb}$ is of second order in
$\lambda_\Delta$ and $\lambda_\Sigma$ because the qubit-mediated
interaction responsible for the effective noise channel in
Eq.~\eqref{eq:effQMEcav} is a second-order effect. This also explains
the, at a first glance, surprising fact that the qubit induces a decay
of the two-resonator system even though its excited state is never
populated. We can understand this by recalling that the
resonator-resonator interaction is mediated not by real, but by
virtual qubit excitations, which are known to give rise to
second-order effects. Equivalently, we may apply a more classical
picture, which is based on the fact that the resonator-resonator
coupling coefficient $g_{\rm SW}$ of Eq.~\eqref{eq:switchmodgs}
depends on the qubit-resonator detuning. Hence, the qubit baths, which
cause first-order fluctuations to the qubit level splitting, induce
second-order fluctuations of $g_{\rm SW}$. The latter are described by
the last term of the master equation~\eqref{eq:effQMEcav}. 

Remarkably, the associated decay rate $\kappa_\mathrm{qb}$ is related
to the qubit relaxation $\gamma$, whereas dephasing $\gamma_\phi$
would enter the effective master equation, Eq.~\eqref{eq:effQMEcav},
only in fourth order in $\lambda_{\{\Delta, \Sigma\}}$ [cf.\
also Eq.~\eqref{eq:effQME}]. Mathematically, this can be understood
from the structure of the dispersive operator
$\sigma_{z,\mathrm{eff}}$ of Eq.~\eqref{eq:sigmazeff}, which couples
the system to dephasing 
noise. To the order $\lambda_{\{\Delta, \Sigma\}}$, this
operator contains products of operators which change the populations
of the qubit and resonators simultaneously. On the one hand, the term
$\sigma^+ a$, for example, describes the excitation of the qubit
together with the emission of a resonator photon, a process which is
energetically forbidden at low temperatures for
$\Delta= \omega_\mathrm{qb}-\Omega> 0$. On the other hand,
terms such as $\sigma^- a^\dagger$ and $\sigma^- b^\dagger$ have no
effect when the qubit remains in the ground state. By contrast, the
operator $\sigma_{x,\mathrm{eff}}$ of Eq.~\eqref{eq:sigmaxeff}, which
is responsible for the qubit energy relaxation, contains terms
such as $\sigma_z(a+b)$ of the order $\lambda_{\{\Delta,
 \Sigma\}}$. These describe a resonator decay without exciting 
the qubit, which is energetically favorable at low temperatures. For
this reason, the only remaining contribution to qubit-enhanced decay
up to second order in $\lambda_{\{\Delta, \Sigma\}}$ in the
effective quantum master equation for the two resonators,
Eq.~\eqref{eq:effQMEcav}, stems from qubit relaxation. The
fourth-order contribution to the dephasing is related to the
appearance of corresponding operators of the order 
$\lambda_{\{\Delta, \Sigma\}}^2$
in $\sigma_{z,\mathrm{eff}}$, which change the
states of the resonators but not that of the qubit.

% *** SECTION ***
\section{Numerical results\label{sec:results}}
We now investigate the validity of the effective
Hamiltonian~\eqref{Hdiag1} with respect to the resonator-resonator
coupling constant, Eq.~\eqref{eq:switchmod} and the effective quantum
master equation~\eqref{eq:effQMEcav} for the resonators. Therefore, we
compare the analytical results derived in the previous sections to
numerically exact results obtained with the Bloch-Redfield quantum
master equation, Eq.~\eqref{eq:blochredfield} using the full
Hamiltonian~\eqref{eq:HS}. For further convenience, we 
assume uniform resonator decay rates,
$\kappa= \kappa_A= \kappa_B$. In our numerical simulations
we use conservative estimates for the qubit decay rates. This is to
stress the effect of the qubit dissipation channel on the resonators.
\begin{figure}
  \centering
  \includegraphics[scale=.9]{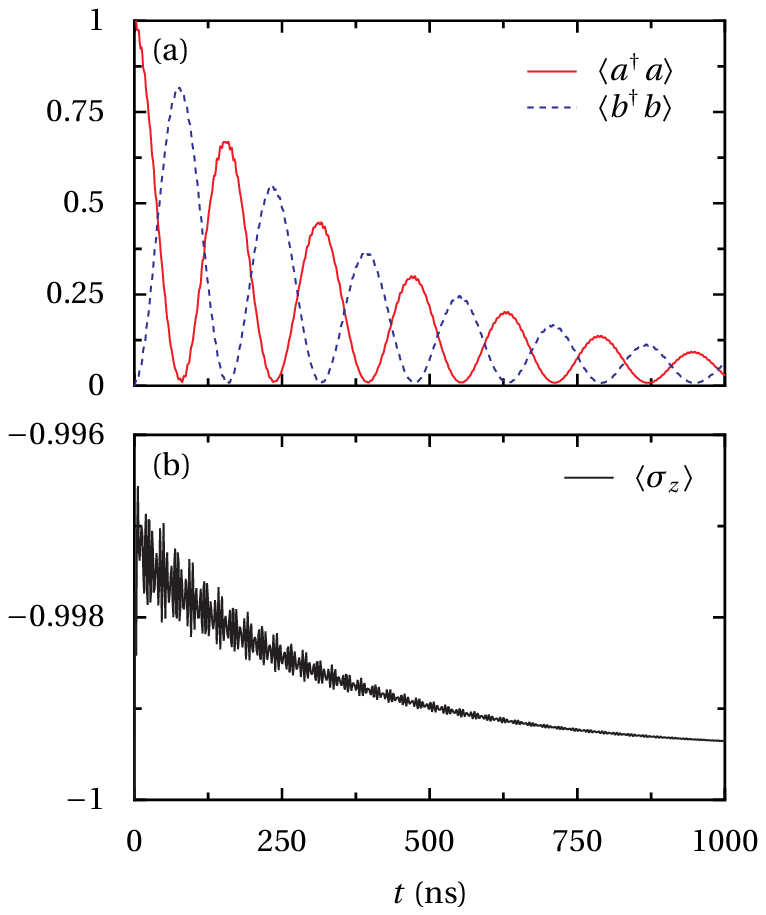}
\caption{(Color online) (a) Rabi oscillations between the 
resonators at the effective interaction strength $g_\mathrm{GSW}/2
\pi= 0.00315\,\text{GHz}$ (numerical value, analytically we find
$g_\mathrm{GSW}/2 \pi= 0.00328\,\text{GHz}$): Occupation numbers of
resonator $A$ (solid red lines) and $B$ (dashed blue lines).
The initial state is ${|} g \rangle _\mathrm{qb} {|} 1 \rangle _A {|}
0 \rangle_B$. (b) Time 
evolution of the qubit state in terms 
of $\langle \sigma_z\rangle$ (solid black lines). The deviation of 
the ideal case $\langle \sigma_z\rangle = {-}1$ always lies 
below  $0.5\%$.  Numerical parameters: 
$\Omega/2\pi = 3.5\, \text{GHz}$, $\delta_Q/ 2 \pi =  4\,
\text{GHz}$, $\varepsilon /2 \pi = {-} 6.37\, \text{GHz} $, 
$g/ 2 \pi =   0.24\, \text{GHz}$, $G/ 2 \pi = 0.0022\,
\text{GHz}$, $T  = 20\, \mathrm{mK}$. Decay 
rates: $J_A (\Omega)/2 \pi =  J_B (\Omega) /2 \pi =  
0.00035\, \text{GHz} $, $J_x (\Omega) /2 \pi = J_z (\Omega)/2
\pi= 0.035 \,\text{GHz}$.} 
\label{fig:Rabi}
\end{figure}

\subsection{Rabi oscillations\label{sec:rabi}}
The observation of Rabi oscillations between the two resonators is a
first feasible application to probe the two-resonator setup. 
A system prepared in the product state ${|} g \rangle
_\mathrm{qb} {|} 1 \rangle _A {|} 0 \rangle_B$ is subject to a
periodic exchange of the excitation between the resonators as long as
their coupling is finite, $g_\mathrm{SW} \neq 0$. The corresponding
oscillation time is $T_\mathrm{Rabi}= \pi/g_\mathrm{SW}$. The initial
excitation could be provided to one of the resonators by means of an
ancilla qubit. For this purpose, suitable protocols have recently been
proposed.~\cite{Hofheinz2008a,Hofheinz2009a,Johansson2006b,Liu2004a}

Figure~\ref{fig:Rabi}(a) depicts the according behavior of the
resonator populations as a function of time. We find the numerically 
observed oscillation period to be in good agreement to
$T_\mathrm{Rabi}$. Note that we have already incorporated the effects
of the dissipative environments 
modeled by the Bloch-Redfield master equation,
Eq.~\eqref{eq:blochredfield} (see discussions in the following
sections). The time evolution of the qubit population 
$\langle \sigma_z\rangle$ is plotted in Fig.~\ref{fig:Rabi}(b). 
From this we can verify that the qubit remains in its ground state
after weak initial transients. These findings 
substantiate the validity of the effective Hamiltonian \eqref{Hdiag1}
in the dispersive regime.

\subsection{Decay rates \label{sec:decayrates}}

In the following we are interested in
understanding quantitatively the influence of the reservoirs on the
two-resonator setup. For this purpose we first make an analytical
estimation based on the effective quantum master
equation~\eqref{eq:effQMEcav}, which are compared then to numerical
results obtained with Eq.~\eqref{eq:blochredfield}. 
We investigate the time evolution of particular observables, the
associated operators of which are constants of motion with respect to 
the dynamics of the closed system. Thus, any dynamics is produced by
the dissipators of the quantum master equations
\eqref{eq:blochredfield} or \eqref{eq:effQMEcav}, respectively. 

\begin{figure}
  \centering
  \includegraphics[scale=.9]{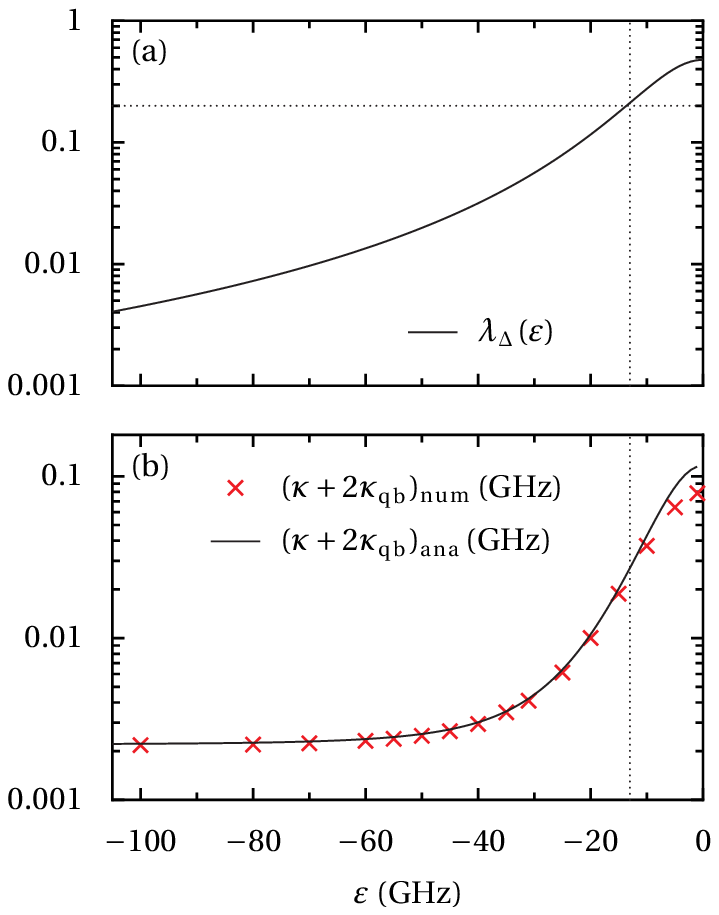}
  \caption{(Color online) Dependence of the parameter
  $\lambda_\Delta$ (a) and the effective damping rate
  $\kappa+2\kappa_\mathrm{qb}$ (b) (black solid lines), on
  the qubit energy bias $\varepsilon$. The rate for  qubit-enhanced decay
  $\kappa_\mathrm{qb}$ is given by Eq.~\eqref{eq:gammaeff}. The
  numerically extracted damping rates (red crosses in (b)) are related
  to the decay of the expectation value of the number operator
  corresponding to the ``center of mass'', $\langle A_+ ^{\dagger}  
  A_+ \rangle$, out of the initial 
  state  ${|} g\rangle_\mathrm{qb} {|}1 \rangle _A {|}0 \rangle_B$  at
  different $\varepsilon$. The dotted lines mark the limit of validity
  of the dispersive theory (see text). Parameters are chosen as in
  Fig.~\ref{fig:Rabi}.}   
\label{fig:qmerate} 
\end{figure}

At this point we recall that the effective Hamiltonian simply
describes a set of two coupled harmonic oscillators as long as the
qubit remains in its ground state. They are each coupled to
independent noise channels, as well as to a joint channel of 
qubit-induced correlated noise via their ``center of mass
coordinate''. The latter is defined as the bosonic operator 
$A_{+}= (a+b)/\sqrt{2}$. In addition we define the ``relative
coordinate'' $A_{-}= (a-b)/\sqrt{2}$. In terms of these 
normal modes the oscillators are not coupled. The associated 
number operators are constants of motion, $[\mathcal H_\mathrm{eff},
A_{\pm}^{\dagger} A_{\pm}^{\phantom{\dagger}}]= 0$. For further
progress we compute the time evolution of the averages $\langle
A_{\pm}^{\dagger} A_{\pm}^{\phantom{\dagger}} \rangle$ using the
effective quantum master equation for the two-resonator system
\eqref{eq:effQMEcav}. Here we note that the evolution of any
operator $\mathcal O$ without explicit time dependence is described
by the adjoint of the quantum master equation~\eqref{eq:effQMEcav},
\begin{align}
  \label{eq:effQMEcavadj}
 \partial_{t} \langle \mathcal O\rangle &=
\frac{i}{\hbar} \Big\langle \Big[\mathcal H ^\mathrm{eff}
_\mathrm{cav},\mathcal O  \Big] \Big\rangle  \\
\nonumber &+  \kappa \big\langle D^{\dagger}
[ a ] \mathcal O \big\rangle + \kappa \big\langle
D ^{\dagger} [ b ] \mathcal O \big\rangle
+  \kappa_\mathrm{qb}
\big \langle D^{\dagger} [a+b] \mathcal O  \big \rangle\, ,  
\end{align}
with
\begin{equation*}
\partial_{t} \langle \mathcal O \rangle 
= \rm {tr} (\partial_{t} \rho \, \mathcal O) \, .
\end{equation*}
The adjoint Lindblad super-operators $D^{\dagger}$ act on the 
operator $\mathcal O$, according to 
\begin{equation}
\label{eq:adjLindblad}
D^{\dagger} [X] \mathcal O = X^{\dagger} \mathcal O X -
\frac{1}{2}[X^{\dagger} X, \mathcal O]_{+} \, ,
\end{equation}
cf. Eq.~\eqref{eq:dissipator}.
Evaluating this relation for the normal mode number operators
$\mathcal O = A_{\pm}^{\dagger} A_{\pm}^{\phantom{\dagger}}$ yields
\begin{equation}
\label{eq:decayplus}
 \langle A_{+}^\dagger A_{+}^{\phantom{\dagger}} \rangle (t) =
 \langle A_{+}^\dagger A_{+}^{\phantom{\dagger}}  \rangle_{t=0} \,
 e^{-(\kappa + 2 \kappa_\mathrm{qb} )t}  
\end{equation}
and
\begin{equation}
\label{eq:decayminus}
 \langle A_{-}^\dagger A_{-}^{\phantom{\dagger}} \rangle (t) =
 \langle A_- ^\dagger A_- ^{\phantom{\dagger}}  \rangle_{t=0}  \,
  e^{-\kappa t} \, .  
\end{equation}
The normal modes are thus expected to decay exponentially. Remarkably,
these decays should occur at different rates. The qubit-induced noise
channel only couples to the center of mass, which suffers
enhanced decay. This becomes manifest in the contribution $2
\kappa_\mathrm{qb}$ 
to the exponent in Eq.~\eqref{eq:decayplus}, with the rate
$\kappa_\mathrm{qb}$ from Eq.~\eqref{eq:gammaeff}. 
The relative coordinate is not affected by the qubit noise
channel, however, and simply decays with the resonator decay rate
$\kappa$, see Eq.~\eqref{eq:decayminus}. Formally, this is because of 
\begin{figure}
  \centering
  \includegraphics[scale=.9]{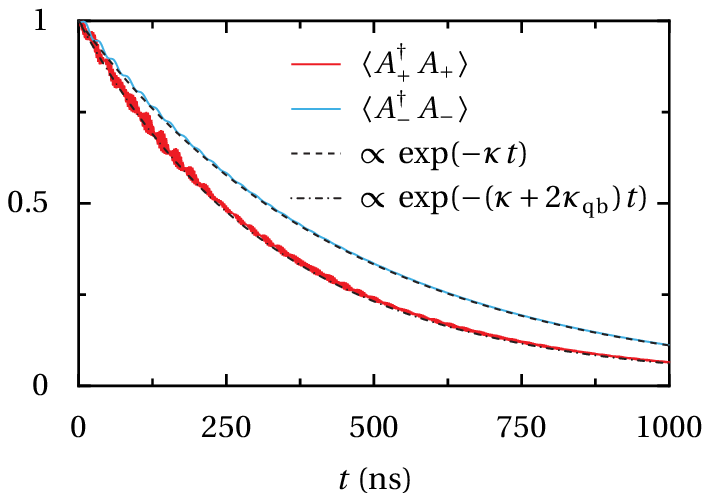}
  \caption{(Color online) Decay of the quantum switch out of the
    initial state ${|} g \rangle _\mathrm{qb} {|} 1 \rangle _A {|} 0
    \rangle_B$. Numerical data obtained with the Bloch-Redfield QME 
    (\ref{eq:blochredfield}) and $N= 3$ states in each resonator:  
    Decay of the expectation values of the number operators
    corresponding to the  ``center of mass coordinate'', $\langle A_+ 
    ^\dagger A_+\rangle$ (red solid lines) and ``relative coordinate'', 
    $\langle A_-^\dagger A_- \rangle$ (blue solid lines), compared to
    analytical estimates, Eqs.~\eqref{eq:decayplus} and
    \eqref{eq:decayminus}  (black dashed and dash-dotted lines,
    respectively). Decay rates are according to
    Eq.~\eqref{eq:effQMEcav}. Parameters: see
    Fig.~\ref{fig:Rabi}.}\label{fig:cavdecay-fock} 
\end{figure}
 \begin{equation*}
 D ^{\dagger} [a+b ]   A_{-}^{\dagger} A_{-}
 ^{\phantom{\dagger}}   = 0 \, . 
 \end{equation*} 
In order to test these analytical estimations based on 
Eq.~\eqref{eq:effQMEcav}, we consider a decay scenario with the
resonators initially prepared in the Fock states ${|} 1
\rangle_A$ (resonator $A$) and ${|} 0 \rangle _B$ (resonator
$B$). The qubit is prepared in its ground state ${|} g \rangle
_\mathrm{qb}$. We calculate numerically the time evolution of the
number operators related to the ``center of mass'', $\langle
A_{+}^\dagger A_{+}^{\phantom{\dagger}} \rangle $, and the ``relative
coordinate'', $\langle A_{-}^\dagger
A_{-}^{\phantom{\dagger}}\rangle$, and compare the 
decay characteristics to the ones suggested by
Eqs.~\eqref{eq:decayplus} and \eqref{eq:decayminus}, respectively. The
results are depicted in Fig.~\ref{fig:cavdecay-fock} for a particular
set of parameters. We find an excellent agreement of theory and
numerical data. While $\langle A_{-}^\dagger A_{-}^{\phantom{\dagger}}
\rangle $ decays at a rate $\kappa$, the decay of $\langle
A_{+}^\dagger A_{+}^{\phantom{\dagger}} \rangle$ is enhanced by the
qubit noise channel, resulting in a decay rate $\kappa +  2
\kappa_\mathrm{qb}$. The latter finding is affirmed in
Fig.~\ref{fig:qmerate}(b), where we compare the analytical expression
for the decay rate $\kappa + 2 \kappa_\mathrm{qb}$ to corresponding
numerical values that are extracted from simulations of a decay
scenario according to Eq.~\eqref{eq:decayplus}. The qubit-induced
decay rate $\kappa_\mathrm{qb}$ is given by
Eq.~\eqref{eq:gammaeff}. Here, we have chosen an explicit dependence 
on the qubit bias energy $\varepsilon$, which is adjustable in
realistic experimental scenarios, while all other parameters are
usually fixed. As indicated by Fig.~\ref{fig:qmerate}(a), the
dispersive description can be considered as valid for
$\lambda_{\{\Delta,\Sigma,\Omega\}}\lesssim 0.1$. These findings
suggest that the effective quantum master equation for the two coupled
resonators \eqref{eq:effQMEcav} describes the dissipative system
behavior adequately in the dispersive limit.  

Furthermore, the stationary value of $\langle A_+^\dagger
A_+^{\vphantom{\dagger}} \rangle$ is found to be different from
zero. This stems from a static energetic shift of the oscillator
potential minima due to a small ``effective force''. The latter
arises from the resonator-qubit coupling component $ \Omega
\lambda_\Omega (a + a^\dagger) \sigma_z$ in the full system
Hamiltonian \eqref{eq:HS}, which has been eliminated in the effective
Hamiltonian \eqref{eq:effQME} by the transformation~\eqref{eq:Z-}. 
The dependence between the original and transformed oscillator
creators and annihilators can be expressed as $a ^{(\dagger)}
\rightarrow a ^{(\dagger)} - \lambda_\Omega$. 
First of all, this shift explains strong oscillations with frequency
$\Omega$ during the time evolution of the resonator states. In
Fig.~\ref{fig:cavdecay-fock}, their effect
is visible, however they are not resolved.
The equilibrium value of the resonator ``center of mass''
population is shifted according to 
\begin{equation}
  \label{eq:staticshift}
  \langle A_+^\dagger A_+^{\vphantom{\dagger}} \rangle  \rightarrow
  \langle A_+^\dagger A_+^{\vphantom{\dagger}} 
  \rangle - 4  \lambda_\Omega^2  \, .
\end{equation}
Following  the same reasoning, the equilibrium value of $\langle
A_-^\dagger A_-^{\vphantom{\dagger}} \rangle$ is zero. 
\begin{figure}
  \centering\includegraphics[scale=.9]{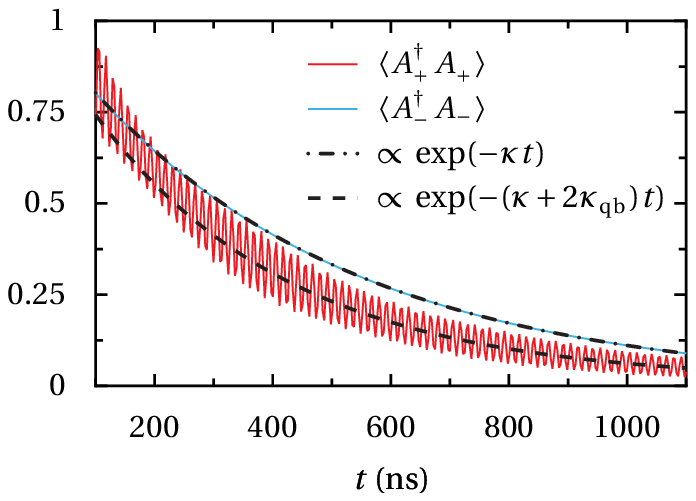} 
  \caption{(Color online) Decay of the quantum switch out of the
    initial coherent  
    state ${|}\alpha \rangle$ with $\alpha= 1$. Simulations were
    run with $N= 6$ states in each resonator. Numerical data:
    occupations of the ``center
    of mass coordinate'', $\langle A_+ ^\dagger A_+\rangle$ (red solid
    lines, oscillating curve)
    and ``relative coordinate'', $\langle A_- ^\dagger A_- \rangle$
    (blue solid smooth line), compared to analytical estimates,
    Eqs.~\eqref{eq:decayplus} and  \eqref{eq:decayminus}  (black
    dashed and dash-dotted lines, respectively). Initial transient
    effects are  not depicted in the figure. Parameters: see
    Fig.~\ref{fig:Rabi}.} 
  \label{fig:cavdecay-coh}
\end{figure}

As a second example, we consider the case of each resonator prepared in a
coherent or Glauber state,
 $ |\alpha \rangle = e^{-|\alpha|^2} \sum_{n=0}^\infty
  \frac{\alpha^n}{\sqrt{n!}} |n\rangle \, $
with $|\alpha|^2$ being the average photon number in the resonator.
This scenario is mainly motivated by experiment, where a coherent
state in a resonator can easily be prepared via a resonant drive. 
To investigate the decay behavior of the ``center of mass''
and ``relative coordinate'' for this scenario, we choose the initial
state ${|} g \rangle _\mathrm{qb} {|} \alpha= 1 \rangle _A {|} \beta =
0  \rangle_B$. As depicted in Fig.~\ref{fig:cavdecay-coh}, the
predictions of the effective quantum master equation are again found
to be in good agreement with our numerical simulations, apart from
transient effects. 

\subsection{Decay of entanglement\label{sec:decayent}}
The generation of entangled two-resonator states is a key application of
the quantum switch. For this purpose, we recall the switching
property of the two-resonator setup mentioned in
Sec.~\ref{sec:beyondRWA}, that is, the possibility to switch on and off
the effective coupling between the resonators by balancing the coupling
coefficient $g_\mathrm{SW}$ given in Eq.~\eqref{eq:switchmod}. While a
similar approach to create entanglement between two resonators based on 
Landau-Zener sweeps \cite{Zueco2008a} has been previously discussed in
Ref.~\onlinecite{Zueco2009b}, we focus on the following,
suitable protocol: A finite interaction strength
$g^\mathrm{on}_\mathrm{SW}$ is initialized by tuning the qubit energy
flux appropriately. After preparing the initial product state ${|} g 
\rangle _\mathrm{qb} {|} 1 \rangle _A {|} 0 \rangle_B$ the two-resonator
state ${|}\psi{\rangle} _\mathrm{cav}$ evolves according to 
\begin{equation}
 \label{eq:evolution}
 |\psi (t) \rangle _\mathrm{cav}
 = \cos (g^\mathrm{on}_\mathrm{SW} t) 
 | 1 \rangle _A | 0 \rangle_B 
 + i  \sin (g^\mathrm{on}_\mathrm{SW} t) 
 | 0 \rangle _A | 1  \rangle_B  \; .
\end{equation}
After a time $t= T_\mathrm{on}$ has elapsed, $g_\mathrm{SW}$ is
balanced back to zero. During the whole procedure the qubit
remains in its ground state ${|} g \rangle_\mathrm{qb}$ and does not
get entangled with the resonators. In particular, 
 $T_\mathrm{on}= \pi/ 4 g_\mathrm{SW}$
results in the entangled two-resonator state $| \psi _+^i\rangle
_\mathrm{cav} = ( {|} 1 \rangle _A {|} 0 \rangle_B + i {|} 0
\rangle _A {|} 1 \rangle_B )/\sqrt{2} $, whereas $T_\mathrm{on} 
=  3\pi/4 g^\mathrm{on}_\mathrm{SW}$ yields the state $|
\psi _-^i \rangle _\mathrm{cav} =  ( {|} 1 \rangle  _A  
{|} 0 \rangle_B - i {|} 0 \rangle _A {|} 1 \rangle_B )/\sqrt{2}
$. A photon transfer from one resonator to the other is accomplished with
$T_\mathrm{on} = \pi/2 g^\mathrm{on}_\mathrm{SW}$. 
\begin{figure}
 \centering
 \includegraphics[scale=.9]{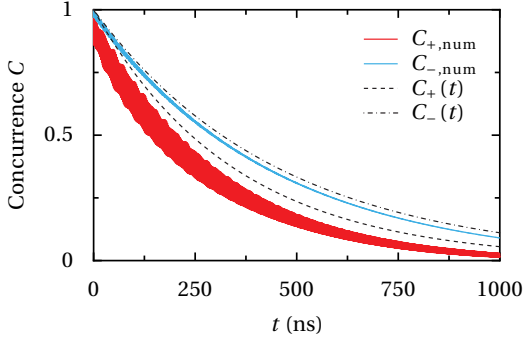}
 \caption{(Color online) Decay of concurrence $C_+$ and $C_-$ for the
   entangled  initial states $|\Psi_+\rangle$ and $|\Psi_-\rangle$,
   (solid red lines, thick curve, and solid blue lines, thin curve), 
   respectively. The switch setting off-condition $g_\mathrm{SW}= 0$
   is fulfilled by $\varepsilon /2 \pi= -8.916\,\text{GHz}/2\pi$.  
   Here we use $G= 0.00478\,\text{GHz}$, all other parameters are as
   in Fig.~\ref{fig:Rabi}. The exponential decay corresponds to
   Eqs.~\eqref{eq:Cplus} and \eqref{eq:Cminus} (black dashed and
   dash-dotted lines,  respectively).}    
 \label{fig:C-alldecay}
\end{figure}

In the above discussions we have disregarded decoherence for reasons
of clarification. In realistic scenarios, however, 
dissipation and dephasing are present even in the case of short
times $T_\mathrm{on}$, which prevents the creation of perfectly
entangled states according to the above described protocol.
Beyond that, two-resonator entanglement once created, will decay with
time, according to the effective two-resonator QME
\eqref{eq:effQMEcav}. In this context, excess phase noise
  in the resonators may cause further adverse effects, which are not
  considered here (cf. Sec.~\ref{sec:systembath}).
Thus, it is important to reveal the decay characteristics of
particular entangled states that could be created in the two-resonator
setup up to a good degree via specific switch setting protocols. For
this purpose, we first focus on the decay characteristics of the
initial entangled two-resonator Bell states 
\begin{equation}
|\psi _\pm \rangle _\mathrm{cav} = \frac{1}{\sqrt{2} }
\big( {|} 1 \rangle _A {|} 0 \rangle_B \pm {|}
0 \rangle _A {|} 1 \rangle_B  \big) \, .\label{eq:bellp}  
\end{equation}
To quantify the entanglement we first assume that all dynamics is
restricted to the subspace $\{|00\rangle, |01\rangle, |10\rangle,
|11\rangle \}$. Thus, we face the dynamics of entanglement between
two two-level systems. In this case, the concurrence $C$ represents 
an adequate measure of entanglement, given by~\cite{Wootters1998a}
$C=\max\{\xi_1-\xi_2-\xi_3-\xi_4,0\}$. The parameters $\xi_j$
denote the ordered square roots of the eigenvalues of the matrix
$\rho_\mathrm{cav} (\sigma_{y}^A  \sigma_{y}^B)
\rho^*_\mathrm{cav} (\sigma_{y}^A \sigma_{y}^B)$ with
$\rho_\mathrm{cav}$ being the reduced density matrix of the two-resonator
state, and $A$ and $B$ labeling the respective resonator Hilbert spaces. 
This representation of the concurrence is quite general and suitable
for numerical investigation. However, for the initial states $|\psi
_\pm \rangle _\mathrm{cav}$ and linear
superpositions hereof, one can obtain analytical expressions for the 
decay characteristics of the concurrence with the help of the
effective quantum master equation \eqref{eq:effQMEcav}. Since the only
nonzero elements of the associated density matrices during the whole
time evolution are $\rho_{00}, \rho_{11}, \rho_{22}$ and $\rho_{12},
\rho_{21}$ in the basis $\{|00\rangle, |01\rangle, |10\rangle,
|11\rangle \}$ the concurrence is simply given by
\begin{equation}
 \label{eq:concurrence}
 C (t) = 2 |\rho_{12} (t)| \; .
\end{equation}
It turns out that the decay characteristics of the density matrix
element $\rho_{12}$ depend on the initial two-resonator state. In
particular, the time evolution of the concurrences $C_\pm (t) $ for the
initial density operators $(|\psi _\pm \rangle \langle \psi _\pm {|}) 
_\mathrm{cav}$ is found as
\begin{align}
C_{+} (t) 
&= e^{-(\kappa + 2 \kappa_\mathrm{qb} ) t} 
\label{eq:Cplus} \\ 
C_{-} (t) 
&= e^{-\kappa t} \label{eq:Cminus} \; .
\end{align}
The reason for this particular behavior is that the state
$ |\psi _-\rangle_\mathrm{cav}$ lies in a decoherence-free subspace
with respect to the dissipator $D[a+ b]$. Thus, it is a robust state
in the sense that it does not couple to the qubit-induced correlated
noise source.~\cite{Doll2006a,Doll2007a} This statement is equivalent
to the relation $D[a + b] (|\psi _- \rangle \langle \psi_-
|)_\mathrm{cav} =0$. On the contrary, the initial state $ |\psi 
_+\rangle_\mathrm{cav}$ is fragile in this respect, since 
$D[a\!+\!b] (|\psi _+ \rangle \langle \psi_+ |)_\mathrm{cav}  \neq
 0$. 

In Fig.~\ref{fig:C-alldecay}, we compare the numerically calculated
time evolution of the concurrence to the analytical results 
of Eqs.~\eqref{eq:Cplus} and \eqref{eq:Cminus}, finding good
agreement. While the decay of $C_+ (t)$ is enhanced due to the qubit
dissipation channel, the time evolution of $C_- (t)$ is
determined by 
resonator dissipation only (cf. Fig.~\ref{fig:C-alldecay}), 
in analogy to the findings of Eqs.~\eqref{eq:decayplus} and
\eqref{eq:decayminus}. We note that a corresponding behavior has been
reported for correlated states of a chain of coupled qubits
interacting with a common bath.~\cite{Ojanen2007a} The numerical
result is found to be shifted with respect to the analytical curves,
since other elements of the density operator, e.\,g. $\rho_{33}$
become populated as well during the time evolution of the system
state. This non-RWA feature stems from the full numerical treatment
using the system Hamiltonian Eq.~\eqref{eq:HSphys}. 
\begin{figure}
 \centering
 \includegraphics[scale=.9]{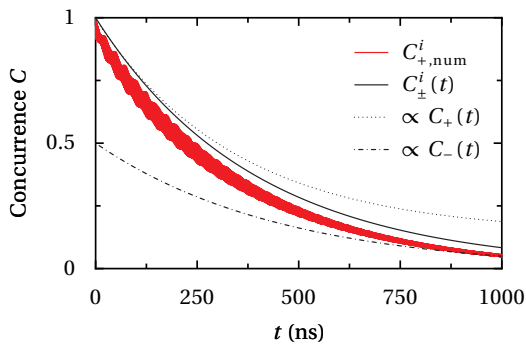}
 \caption{(Color online) 
 Decay of concurrence $C^i _+$ for the entangled initial state
 $|\Psi_+ ^i\rangle$. The switch setting off-condition $g_\mathrm{SW}=
 0$ is fulfilled by $\varepsilon /2 \pi= -8.916\,\text{GHz}$. Here we
 use $G= 0.00478\,\text{GHz}$, all other parameters as in 
 Fig.~\ref{fig:Rabi}. The two-mode exponential (solid
 black line) corresponds to  Eq.~\eqref{eq:CplusI}. The additional 
 exponentials are given by Eqs.~\eqref{eq:Cplus} and
 \eqref{eq:Cminus} and highlight the different decay regimes.} 
 \label{fig:CplusI-alldiss} 
\end{figure}

These findings can now be employed to characterize the decay of
entanglement for the initial states $|\psi_\pm ^i
\rangle_\mathrm{cav}$. For this purpose we express them as linear
superpositions of the Bell states $|\psi_\pm\rangle_\mathrm{cav}$
[Eq.~\eqref{eq:bellp}],  
\begin{equation} \label{eq:psipmI}
 \begin{split}
   |\psi_\pm ^i \rangle _\mathrm{cav}
   =  \frac{1}{2} \Big( (1+i) |\psi_+\rangle + (1\pm i)
   |\psi_-\rangle \Big)_\mathrm{cav} \; .
 \end{split}
\end{equation}
Consistently, we find that the analogously defined concurrences
$C^i_\pm$ can be expressed as a sum of the concurrences
of the initial Bell states [Eqs.~\eqref{eq:Cplus} and
\eqref{eq:Cminus}] as
\begin{equation}\label{eq:CplusI}
C_+ ^i (t) = C_- ^i (t) = \frac{1}{2}\left( e^{-(\kappa + 2
    \kappa_\mathrm{qb} ) t} + e^{-\kappa t} \right) \; .
\end{equation}
This has some interesting consequences. For short times, the decay out
of both initial states $| \psi _\pm^i\rangle_\mathrm{cav}$ is merely
governed by qubit-enhanced decay at a rate $\kappa + 2
\kappa_\mathrm{qb}$. In the limit of long times, however, one finds 
pure resonator decay at a rate $\kappa$. We have confirmed this
numerically in Fig.~\ref{fig:CplusI-alldiss} by means of the
concurrence $C_+ ^i (t)$ related to the initial state $| \psi
_+^i\rangle_\mathrm{cav}$.

In summary, we point out that it is possible to understand the
time evolution 
characteristics of the entanglement in the system on the basis of the 
effective master equation~\eqref{eq:effQMEcav}. We emphasize that the  
qubit-induced dissipation channel plays a crucial, selective role for  
different classes of initially entangled states.
\begin{figure}
  \centering
\includegraphics[scale=.9]{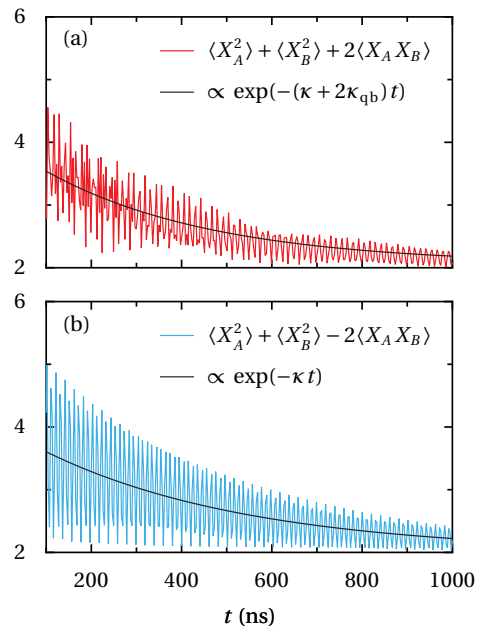}
  \caption{(Color online) Time evolution of the auto- and
    cross-correlations of the initial state  ${|} g \rangle
    _\mathrm{qb} {|} \alpha= 1\,\rangle _A {|} \beta= 0 \rangle_B$:
    Decay of $\langle X_A^2 \rangle + \langle X_B^2 \rangle + 2
    \langle X_A X_B \rangle $ (upper part, red lines) and $\langle
    X_A^2 \rangle + \langle X_B^2 \rangle - 2 \langle X_A X_B \rangle
    $ (lower part, blue lines). The exponential decay is compared to
    the analytical  
    estimates given by \eqref{eq:decayplus} and \eqref{eq:decayminus}
    (black solid lines). Initial transient effects are not depicted in
    the figure. For parameters and initial conditions, see
    Fig.~\ref{fig:Rabi} and Fig.~\ref{fig:cavdecay-coh},
    respectively.}   
  \label{fig:correlations}
\end{figure}

\section{Extracting damping constants by (cross-) %
  correlations\label{sec:corr}}

In the two resonator setup, it is possible to measure correlations and 
cross-correlations in terms of the expectation values 
$\langle (a+a^{\dagger})^{2} \rangle$, $\langle (b+b^{\dagger})^{2}
\rangle$ and $\langle(a+a^{\dagger}) (b+b^{\dagger})\rangle$ 
with present techniques.~\cite{Mariantoni2006a}
In the following we propose a method how to extract the relaxation
rates $\kappa$ and $\kappa _\mathrm{qb}$ out of correlation
measurements of such type. We define the oscillator ``positions''
$X_A= a+ a ^\dagger$ and $X_B= b+ b ^\dagger$. In analogy to
Sec.~\ref{sec:decayrates}, we find analytically that the quantity
$\langle X^2_{-}\rangle = \langle X_A^2 \rangle + \langle X_B^2
\rangle - 2 \langle X_A X_B \rangle $ decays at the rate $\kappa$. For  
$\langle X^2_+ \rangle= \langle X_A^2 \rangle + \langle X_B^2
\rangle + 2 \langle X_A X_B \rangle $ we find a decay with $\kappa +
2\kappa _\mathrm{qb} $. In Fig.~\ref{fig:correlations}, we numerically
substantiate these findings for the example of a coherent initial
state ${|} g \rangle _\mathrm{qb} {|} \alpha = 1 \rangle _A {|}
\beta = 0 \rangle_B$, discarding again transient effects. 

Thus, we point out the possibility to extract the resonator 
decay rate $\kappa$ by measuring the decay of $\langle
X^2_-\rangle$. This allows in turn for deducing the rate $\kappa
_\mathrm{qb}$ related to qubit-enhanced decay by measuring
the quantity $\langle X^2_+\rangle$. From the latter measurement,
it is further possible to gain information about the relaxation and
dephasing rates of the qubit, $\gamma (\omega_\mathrm{qb})$
[Eq.~\eqref{eq:relaxationrate2}] and $\gamma_\phi (\omega\rightarrow
0)$, [Eq.~\eqref{eq:dephasingrate2}], provided that the system
frequencies and resonator-qubit interaction strengths are known. More
details about a possible experimental realization of such correlation
measurements can be found in
Refs.~\onlinecite{Menzel2010a} and~\onlinecite{Mariantoni2010a}.

\section{Conclusions\label{sec:conclusion}}

We have investigated a two-resonator circuit QED setup. In the
dispersive regime, i.\,e. if the resonator-qubit detuning is 
much larger than their mutual coupling, it is possible to extract the
relevant system dynamics by applying the 
unitary transformation \eqref{eq:Schrieffer1} to the system
Hamiltonian \eqref{eq:HS}. The resulting effective Hamiltonian
\eqref{Hdiag1} reveals that the qubit gives rise to a switchable
coupling between the resonators via virtual excitations. 
This dynamical coupling adds to the direct resonator-resonator
coupling. Balancing both contributions, the resonator-resonator
interaction can be set to zero. Such a qubit-mediated interaction 
provides a physical realization for a quantum switch between the
resonators.

As a principal point, we have focused on the dissipative system
properties that stem from the interaction with different
environments. For weak system-bath coupling, it is possible to cast
the time evolution of the reduced system state into a quantum master
equation of Bloch-Redfield form, Eq.~\eqref{eq:blochredfield}. 
It is usually derived starting from the total microscopic system-bath
Hamiltonian, Eq.~\eqref{eq:Hsbeigen}. Its character being quite
general, it only offers limited analytic insight. To study the 
dissipative dynamics in the dispersive regime, it is preferable to
obtain a more useful, effective analytical description of the
dissipative system dynamics. To this end, we have applied the unitary
transformation Eq.~ \eqref{eq:Schrieffer1} to the total system-bath
Hamiltonian~\eqref{eq:Hsbeigen} and analogously obtained the
transformed, effective system-bath coupling operators. Applying
standard methods, we have derived a rather complex effective quantum
master equation for the system state in the rotated frame. It 
can be simplified, however, assuming low temperatures and 
recalling that the qubit state does not change, and only the dynamics
of the two-resonator system are of relevance. By tracing out the
qubit degrees of freedom one arrives at the Lindblad-type quantum
master equation \eqref{eq:effQMEcav} for the reduced two-resonator
state. 

\begin{figure}
  \centering
  \includegraphics[width=.99\columnwidth]{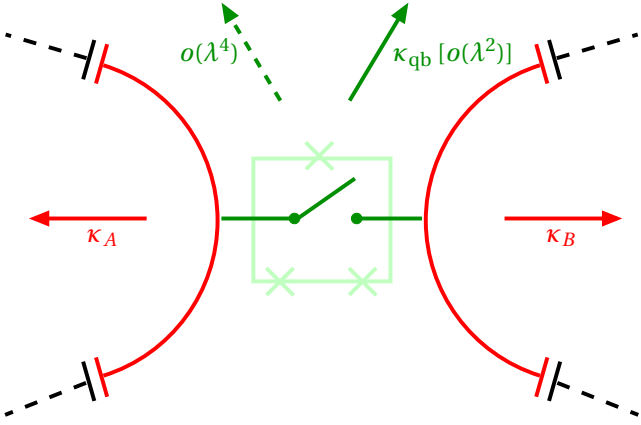}
  \caption{(Color online) Sketch of the relevant decay
   mechanisms affecting the quantum switch. The qubit induces extra
   decoherence in higher dispersive order to the effective
   resonator-resonator dynamics. The arrows mark the decoherence
   channels, labeled by the corresponding decay rates
   $\kappa_{\{A,B,\mathrm{qb}\}}$. Decay processes of fourth order in
   the parameters $\lambda_{\{\Delta,\Sigma,\Omega\}}$ are not
   displayed individually.}      
  \label{fig:two_resonator_eff}
\end{figure}

As a main result of this paper, we have found that the qubit induces
decoherence on the resonator-resonator system via an additional noise
channel that acts on the ``center of mass coordinate'' of the
resonators. This effect stems from qubit energy relaxation and is of
second order in the small dispersive parameters
$\lambda_{\{\Delta,\Sigma\}}$, whereas pure qubit dephasing only
enters in fourth order. This result anticipates that the operation of
the quantum switch is robust against low-frequency noise in the two
level system. For reasons of clearness, our findings are again
illustrated in Fig.~\ref{fig:two_resonator_eff}.
We have verified our analytical findings by numerical calculations,
where we have taken into account the full dissipative dynamics
according to Eq.~\eqref{eq:blochredfield}. As detailed in
Sec.~\ref{sec:results} by means of several examples, we have found an
excellent agreement of the presented dispersive theory and the
numerical results. Here, in particular, we have investigated the
validity of the obtained resonator relaxation rates. With regard to 
generating entangled states, which is a key application of the quantum 
switch, we have examined the decay mechanisms for different
entangled initial resonator-resonator states.

%%%%%%%%%%%%%%%%%% aknowledgments %%%%%%%%%%%%%%%%%%%%%%%%%%%

\begin{acknowledgments}
We gratefully acknowledge financial support by the German Excellence
Initiative via the ``Nanosystems Initiative Munich (NIM)''.
E.S. acknowledges funding from UPV/EHU Grant GIU07/40, Ministerio de
Ciencia e Innovaci\'on FIS2009-12773-C02-01, European Projects
EuroSQIP and SOLID. 
This work has been supported by the German Research Foundation (DFG)
through the Collaborative Research Centers SFB 484 and SFB 631. 
\end{acknowledgments}

% *** APPENDIX ***
\appendix
%
% *** APPENDIX SECTION ***

\section{Energy relaxation and pure dephasing rates of the qubit}
\label{sec:qubitqme}
In this section, we derive a quantum master equation for the qubit
alone, which allows us to identify the energy relaxation and pure
dephasing rates of the qubit. Considering only a qubit coupled to
individual environments along the $\sigma'_x$ and $\sigma'_z$-axes in 
the laboratory frame, the qubit-bath Hamiltonian reads 
\begin{multline}
  \label{eq:Hqubit}
    \mathcal{H'}_\mathrm{tot,qb}
    =\mathcal{H'}_\mathrm{qb}
    + \sum_{\mu = x,z}  Q_\mu \sum_j c_j ^\mu
    \left(d_{j,\mu}^\dagger +  d_{j,\mu} ^{\vphantom{\dagger}}
    \right) \\ 
    + \sum_{\mu = x,z} \sum_j \hbar \omega_j ^\mu
    \left(d_{j,\mu}^\dagger  d_{j,\mu} ^{\vphantom{\dagger}} +
      \frac{1}{2} \right)    \,,
\end{multline}
where $\mathcal{H'}_\mathrm{qb} = (\hbar\epsilon/2)\sigma'_z +
(\hbar\delta_Q/2)\sigma'_x$. Applying Eq.~\eqref{eq:pauli}, we obtain
the diagonal qubit Hamiltonian
$\mathcal{H}_\mathrm{qb} = (\hbar\omega_\mathrm{qb}/2)\sigma_z$ and also
the qubit-bath Hamiltonian,   
\begin{equation}\label{eq:Hqubit1}
  \begin{split}
    \mathcal{H}_\mathrm{tot,qb}={}
    & \mathcal{H}_\mathrm{qb}
    + \sum_{\mu = x,z} \sum_j \hbar \omega_j ^\mu
    \left(d_{j,\mu}^\dagger d_{j,\mu} ^{\vphantom{\dagger}} +
      \frac{1}{2} \right)    \\
    & {}+(\sin\theta\,\sigma_z+\cos\theta\,\sigma_x)
      \sum_j c_j^x \left(d_{j,x}^\dagger + d_{j,x}
      ^{\vphantom{\dagger}}  \right) \\
    & {}+(\cos\theta\,\sigma_z-\sin\theta\,\sigma_x)
     \sum_j c_j^z
     \left(d_{j,z}^\dagger + d_{j,z} ^{\vphantom{\dagger}} \right)
    \; .
  \end{split}
\end{equation}
Starting from the density matrix $\varrho_\mathrm{tot,qb}$
associated with $\mathcal{H}_\mathrm{tot,qb}$ and following the
lines of Ref.~\onlinecite{Breuer2001a}, the Lindblad quantum master
equation for the reduced qubit density operator
$\varrho_\mathrm{qb} = \mathrm{Tr}_\mathrm{bath}[\varrho_\mathrm{tot,qb}]$
can be derived. To this end, the spectral decompositions
$\sigma_-(\omega) =  \delta(\omega_\mathrm{qb} - \omega)\sigma_-$,
$\sigma_+(\omega) = \delta(\omega_\mathrm{qb} + \omega)\sigma_+$, and
$\sigma_z(\omega) = \delta(\omega)\sigma_z$ of the qubit-bath coupling
operators are required. Omitting the explicit time dependence of
$\varrho_\mathrm{qb}$ for simplicity, we find  
\begin{equation}\label{eq:qme-qubit}
  \begin{split}
    \dot\varrho_\mathrm{qb}={} 
    & {}-\frac{i}{\hbar}\Big[\mathcal{H}_\mathrm{qb}
     ,\varrho _\mathrm{qb}\Big]
    \\
    & {}+\gamma(\omega_\mathrm{qb})
    \Big[\sigma^-\varrho_\mathrm{qb}\sigma^+
    -\frac{1}{2}\big[\sigma^+\sigma^-,\varrho_\mathrm{qb}
    \big]_+\Big]
    \\
    & {}+\gamma(-|\omega_\mathrm{qb}|)
    \Big[\sigma^+\varrho_\mathrm{qb}\sigma^-
     -\frac{1}{2}\big[\sigma^- \sigma^+,\varrho_\mathrm{qb}
     \big]_+\Big]
    \\
    &{}+\gamma_\phi(\omega{\rightarrow}0)
     \Big[\sigma_z\varrho _\mathrm{qb}  
     \sigma_z-\varrho_\mathrm{qb}\Big] \, .
  \end{split}
\end{equation}
Here, $\sigma_- = |g\rangle\langle e|$ and
$\sigma_+ = |e\rangle\langle g|$ are the fermionic qubit annihilation
and creation operators. The energy level transition and the pure
dephasing rates are given by
\begin{align}
  \gamma(\omega)
  & {}=\Gamma_x(\omega)\cos^2\theta+\Gamma_z(\omega)\sin^2\theta
  \label{eq:relaxationrate}
  \\
  \gamma_\phi(\omega)
  & {}=\Gamma_x(\omega)\sin^2\theta+\Gamma_z(\omega)\cos^2\theta \; , 
  \label{eq:dephasingrate}
\end{align}
respectively, and depend on the bath correlation functions
\begin{equation}\label{eq:correlationcases}
 \Gamma_\mu (\omega) = 
\begin{cases}
 J_\mu (\omega) \, (n_\mu(\omega) + 1) \qquad &\omega \geq0 \\
  J_\mu (|\omega|) \, n_\mu(|\omega|)  \qquad &\omega <0 \; ,
\end{cases}
\end{equation}
where $n_\mu(\omega) = 1/(e^{\hbar\omega/k_{\mathrm{B}} T_{\mu}} - 1)$
is the Bose distribution function of bath with label $\mu \in
\{x,z\}$. Because 
the quantum switch operates in the limit of low temperatures,
$k_\mathrm{B}T_\mu \ll \hbar\omega_\mathrm{qb}$, the Bose-factor
$n_\mu(\omega)$ vanishes for frequencies of the order of
$\omega_\mathrm{qb}$. However, for $\omega \rightarrow 0$,
$n_\mu(\omega)$ tends to diverge. This can be relevant in the
experimentally important case of
1/f-noise,~\cite{Yoshihara2006a,Deppe2007a,Kakuyanagi2007a} which
would require a treatment beyond the framework of a Markovian
master equation, exceeding the scope of this work. Instead, we avoid
the divergence problem by choosing Ohmic spectral densities 
[Eq.~\eqref{eq:ohmic}] for low frequencies
$\omega \ll \omega_\mathrm{qb}$. In many cases, this assumption is
reasonable.~\cite{vanderWal2001a,vanderWal2003a,Thorwart2004a}
Provided that both baths have the same temperature
$T= T_{\{x ,z\}}$, we obtain  
\begin{align}
\gamma(\omega\ge0) &= J_x(\omega)\cos^2\theta+J_z(\omega)\sin^2\theta
\label{eq:relaxationrate1} \\
\gamma(\omega<0) & \approx0 
\label{eq:excitaionrate1} \\
\gamma_{\phi}(\omega \rightarrow 0)  &= \frac{k_\mathrm{B}T}{\hbar}
 (\alpha_z\cos^2\theta+\alpha_x\sin^2\theta)\,.
\label{eq:dephasingrate1}
\end{align}
These rates are functions of the mixing angle $\theta$.
Eq.~\eqref{eq:relaxationrate1} constitutes the
main result of this section. For $\omega= \omega_\mathrm{qb}$, it
establishes the connection between the energy relaxation rate in the
energy eigenframe and the dissipative baths defined in the laboratory
frame. The Markovian description of Eq.~\eqref{eq:qme-qubit} remains
justified as long as the spectral densities $J_{\{x,z\}}(\omega)$ are
smooth functions in $\omega_\mathrm{qb}$. This allows one
to apply a linear approximation in an infinitely small interval
around $\omega_\mathrm{qb}$, which yields an effectively Ohmic
description. We finally note that in the special case 
of noise coupling purely via the laboratory $z$-axis (e.g., a flux
qubit exposed to flux noise) the result 
$\gamma(\omega_\mathrm{qb})= J_z(\omega_\mathrm{qb})\sin^2\theta$ is
in agreement with findings from other
works.~\cite{Ithier2005a,Schriefl2006a,Deppe2007a}
For the sake of completeness we mention that the master equation
\eqref{eq:qme-qubit} together with the rates
\eqref{eq:relaxationrate1}, \eqref{eq:excitaionrate1} and
\eqref{eq:dephasingrate1} reproduce the well-known results concerning 
relaxation and pure dephasing times,
$(T_1)^{-1}= \gamma(\omega\ge0)$ and $T_\phi ^{-1} =
\gamma_{\phi}(\omega \rightarrow 0) $ respectively.

\section{Effective bath coupling operators}
\label{sec:effectivecoupling}
To obtain the quantum master equation for the reduced system state, it
is necessary to transform the total system-bath Hamiltonian to the
dispersive picture via the transformation
$\mathcal{H}_{\mathrm{tot,eff}}= \mathcal{U}^\dagger 
\mathcal{H}_\mathrm{tot}\mathcal{U}$, with $\mathcal U$ given in
Eq.~\eqref{eq:Schrieffer1}. The effective system 
Hamiltonian having been derived in Sec.~\ref{sec:beyondRWA}, we need
yet to find the transforms~\eqref{eq:Schrieffer1} of the system-bath
coupling operators $Q_\mu$ to the dispersive frame. Up to 
second order in $\lambda_{\{\Delta,\Sigma,\Omega\}}$, they read as

\begin{equation}
  \begin{split}
    \label{eq:transfcpl1}
    Q _{\mu,\mathrm{eff}} 
    = & \mathcal U^\dagger Q_\mu \mathcal  U  \\
     = & Q_\mu + \big[ Q_\mu, \lambda_\Delta \mathcal{D}  +
     \lambda_\Sigma\mathcal{S} + 
    \lambda_\Omega\mathcal{W} \big] \\
    & +  \Big[ \big[ Q_\mu, \lambda_\Delta \mathcal{D}  +
    \lambda_\Sigma\mathcal{S} + 
    \lambda_\Omega\mathcal{W} \big], 
    \lambda_\Delta \mathcal{D}  +
    \lambda_\Sigma\mathcal{S} + 
    \lambda_\Omega\mathcal{W} \Big] \\
     &+ \mathcal O \big(\lambda_{\{\Delta,\Sigma,\Omega\}}^3\big) 
     \qquad \qquad  \mu \in \left\{ A,B,x,z \right\}   \, .   
  \end{split}
\end{equation}
Each of the effective bath coupling operators $Q_{\mu, \mathrm{eff}}$
is represented by of a sum of operators,   

\begin{equation*}
Q_{\mu, \mathrm{eff}} = \sum_{j_\mu} Q_{j_\mu} \; .
\end{equation*}
For the resonator-bath coupling operators $Q_A= a + a^\dagger$  
and $Q_B = b + b^\dagger$, we obtain the dispersive transforms
as  

\begin{widetext}
  \begin{align}
   \nonumber Q_{A,\mathrm{eff}} = (a + a ^\dag ) _\mathrm{eff} =
    & a + a^\dagger +
    (\lambda_\Delta - \lambda_\Sigma) \sigma_x
    - 2 \lambda_\Omega \sigma_z \\
    & \phantom =  + \frac{1}{2} (\lambda_\Delta^2 - 
     { \lambda_\Sigma} ^2) (a+b + a^\dagger +   b^\dagger)
     + \lambda_\Omega (\lambda_\Delta + \lambda_\Sigma)
    \sigma_x (a+b + a^\dagger + b^\dagger) \, ,
    \label{eq:XAeff}\\[.5ex]
    \nonumber Q_{B,\mathrm{eff}} = (b + b ^\dag ) _\mathrm{eff} =
    &  b + b^\dagger +  (\lambda_\Delta - 
     \lambda_\Sigma)  \sigma_x - 2 \lambda_\Omega \sigma_z \\
     & \phantom =   + \frac{1}{2}
    (\lambda_\Delta^2 - {\lambda_\Sigma}^2) (a+b + a^\dagger + b^\dagger)
     + \lambda_\Omega (\lambda_\Delta + \lambda_\Sigma)
    \sigma_x (a+b + a^\dagger + b^\dagger) \, .
    \label{eq:XBeff}
  \end{align}
\end{widetext}
The dispersive transforms of the qubit-bath coupling operators 
$Q_{x,\mathrm{eff}}= \sigma_{x,\mathrm{eff}}$ and
$Q_{z,\mathrm{eff}}= \sigma_{z,\mathrm{eff}}$ are obtained as a
combination of those of the qubit operators $\sigma'_x$
and $\sigma'_z$ in the laboratory basis, according to
Eq.~\eqref{eq:pauli}. The latter assume the form

\begin{widetext}
\begin{equation}\label{eq:sigmaxeff}
  \begin{split}
    \sigma'_{x, \mathrm{eff}} =
    &  \sigma_x    
    + ( \lambda_\Delta + \lambda_\Sigma ) \sigma_z (a+b + a^\dagger +
    b^\dagger)  
    - 2 \lambda_\Omega \Big[ \sigma^+ (a+b) + \sigma^- (a^\dagger +
    b^\dagger)   - \sigma^+(a^\dagger + b^\dagger) - \sigma^- (a
    + b )\Big] \\
    &-  \Big( (\lambda_\Delta + \lambda_\Sigma )^2 - 4 \lambda_\Omega ^2
    \Big)  \sigma_x  \Big[(a^\dagger  + b^\dagger)(a+b) +1 \Big] 
    + \Big(2 \lambda_\Omega ^2 - (\lambda_\Delta + \lambda_\Sigma)
    \lambda_\Delta \Big) \Big[ \sigma^- (a^\dagger + b^\dagger)^2 + \sigma^+
    (a+b)^2 \Big] \\
    & + \lambda_\Omega (\lambda_\Delta - \lambda_\Sigma)  \sigma_z (a+b +
    a^\dagger + b^\dagger) + \Big(2 \lambda_\Omega ^2 - (\lambda_\Delta +
    \lambda_\Sigma) \lambda_\Sigma\Big) \Big[ \sigma^- (a + b )^2 +
    \sigma^+ (a^\dagger + b^\dagger)^2 \Big] \, ,
  \end{split}
\end{equation}

\begin{equation}\label{eq:sigmazeff}
  \begin{split}
    \sigma'_{z,\mathrm{eff}}  = & \sigma_z  
    - 2 \lambda_\Delta \Big[\sigma^+ (a+b) 
    + \sigma^- (a^\dagger + b^\dagger) \Big]
    - 2 \lambda_\Sigma \Big[ \sigma^+(a^\dagger +
    b^\dagger) + \sigma^- (a + b )\Big] \\
    &  -  2 \sigma_z   \Big(\lambda_\Delta^2 + \lambda_\Sigma ^2\Big)  
    \Big[(a^\dagger +  b^\dagger)(a+b) +1 \Big]  + 4 \lambda_\Omega
    (\lambda_\Delta - \lambda_\Sigma) \sigma_x \Big[ 
    (a^\dagger + b^\dagger)(a+b) +1 \Big]\\
    & + 4 \lambda_\Omega \lambda_\Delta
    \Big[ \sigma^+ (a+b)^2 + \sigma^- (a^\dagger + b^\dagger)^2 \Big] 
     + 4  \lambda_\Omega \lambda_\Sigma \Big[ \sigma^- 
    (a + b )^2 + \sigma^+ (a^\dagger + b^\dagger)^2 \Big] 
    - 2 \lambda_\Delta \lambda_\Sigma \sigma_z \Big[(a+b)^2  
    + (a^\dagger + b^\dagger)^2 \Big]   \, .
  \end{split} 
\end{equation}
\end{widetext}

\section{Effective quantum master equation in the dispersive
  limit \label{sec:deriving}} 

Starting from the Bloch-Redfield quantum master
equation \eqref{eq:blochredfield}, we move to an interaction picture
with respect to the system and the individual reservoirs. Here, the
coupling operators $Q_\mu$ have to be replaced by their dispersive transforms
$Q_{\mu, \mathrm{eff}}$ found in App.~\ref{sec:effectivecoupling}. Now,
we introduce the spectral decompositions
\begin{equation}
  \label{eq:spectral}
  Q_{\mu, \mathrm{eff}} \equiv \sum  _{j_\mu} Q_{j_\mu} = 
\sum _{j_\mu} \sum_\omega Q_{j_\mu}  (\omega)\, .
\end{equation}
The $Q_{j_\mu}$ are the summands of the effective coupling
operators as detailed in Eqs.~\eqref{eq:XAeff}-\eqref{eq:sigmazeff}. 
The spectral components $Q_{j_\mu}(\omega)$ are obtained by
expanding the $Q_{j_\mu}$ in terms of the eigenstates of the
effective Hamiltonian \eqref{Hdiag1}, which we cast 
in entirely diagonal form for this reason,
\begin{equation}
   \mathcal H _\mathrm{eff} 
  =  \hbar \hat {\tilde \Omega}_+ 
  \Big( A_+^\dagger A_+ ^{\vphantom{\dagger}} + \frac{1}{2} \Big)  
  + \hbar \tilde \Omega_- \Big( A_-^\dagger A_- ^{\vphantom{\dagger}} 
  + \frac{1}{2}\Big)  
  +\frac{\hbar \tilde \epsilon}{2} \, ,
 \label{eq:Hdiagn}
\end{equation}
Here we have defined 
\begin{align}
  \hat {\tilde \Omega}_+ & =  \Omega + G 
  + 2 (\lambda_\Delta \Delta 
  + \lambda_\Sigma \Sigma) \sigma_z  \label{eq:newfreq1}\\
  \tilde \Omega_- & = \Omega - G  \label{eq:newfreq2} \\
 \tilde \epsilon &= \omega_\mathrm{qb} 
  + 2 (\lambda_\Delta \Delta + \lambda_\Sigma \Sigma)
 \label{eq:newfreq3} \; ,
\end{align}
and introduced via a linear transformation the normal modes
\begin{equation}
  \label{eq:transform}
  A_+  =\frac{1}{\sqrt{2}} \big (  a + b \big ),  \hspace{1em} 
  A_-  =\frac{1}{\sqrt{2}} \big (  a - b \big ) \, .
\end{equation}
The eigenstates of the effective Hamiltonian (\ref{eq:Hdiagn}) can be
considered as re-defined Fock states,  
\begin{eqnarray}
  \nonumber  \mathcal H _\mathrm{eff} |n  m l\rangle &=& 
  \frac{\hbar \tilde \epsilon}{2} (-1)^{l+1} |l\rangle
  +  \hbar  {\tilde \Omega}_+ (l) \left( n +\frac{1}{2}\right)
  |n\rangle  \\
  &+& \hbar \tilde \Omega_-  \left( m +\frac{1}{2}\right)
  |m\rangle  \, . \label{eq:eigenstates}
\end{eqnarray}
Here,  $\{n,m\}  =  \{0,1,2,\ldots\}$ denote the oscillator
excitations (or resonator photon numbers), and $\tilde \Omega_A (l)$ can 
assume the values  
\begin{equation}
\tilde \Omega_+ (l) = \Omega + G 
+ 2 (\lambda_\Delta \Delta + \lambda_\Sigma \Sigma)
 (-1)^{l+1} \, .
\label{eq:qubitstate}
\end{equation}
with $l= 1$ or $l= 0$ for the qubit being in the excited or
ground state, respectively. With this at hand we find
the spectral decompositions of the effective coupling operators via
the ansatz 
\begin{multline}
  \label{eq:decomposition}
   Q_{j_\mu} (\omega) = \sum_{nml} \sum_{n'm'l'}
   \delta (\Delta^{n'm'l'} _ {nml} - \omega)  \\
 \times |nml\rangle \langle nml|  Q_{j_\mu}  
   |n'm'l'\rangle  \langle n'm'l'| \, ,
\end{multline}
where $\Delta^{n'm'l'} _ {nml}$ denotes the energy difference between
the states ${|}n'm'l'\rangle$ and ${|}nml\rangle$.
For illustration we list the explicit expressions for the
spectral decompositions of some components, 
\begin{align}
  \begin{split}
    A_+ (\omega) &= A_+ |0 \rangle \langle 0 | \, 
    \delta (\omega - \tilde \Omega_+ (0))  \\
    & \quad + A |1 \rangle \langle 1 | \, 
    \delta (\omega - \tilde \Omega_+ (1))
  \end{split} \label{eq:components1} \\
 A_- (\omega) & =  A_- \, \delta (\omega - \tilde \Omega_-)
 \label{eq:components2}\\ 
 \begin{split}
   \sigma^- (\omega) &= \sigma^- \sum_n |n \rangle \langle n | \,
   \delta \Big(\omega - \big[\omega_\mathrm{qb} + (2n+1) \\
   & \quad \times \big(\lambda_\Delta ^2 \Delta + \lambda_\Sigma ^2
   \Sigma \big)\big]\Big) 
 \end{split} \label{eq:components3} \\
\sigma_z (\omega) & =  \sigma_z \, \delta(\omega)
\label{eq:components4} \, .
\end{align}
The decompositions of operator products such as
$\sigma^- A_+^\dagger$ etc. are obtained analogously via the
relation $Q_{j_\mu} ^\dagger (\omega)= Q_{j_\mu}
(-|\omega|)$. With this and Eq.~\eqref{eq:spectral}, we recast the 
Bloch-Redfield quantum master equation \eqref{eq:blochredfield} into
the form   
\begin{multline} \label{eq:blochredfield1}
   \dot \rho (t) = 
   -\frac{i}{\hbar} \Big[ \mathcal H_\mathrm{eff}, 
   \rho (t) \Big]  
   + \sum_{\mu} \sum_{j_\mu,  k_\mu} 
 \sum_{\omega,  \omega'} e^{i(\omega'-\omega)t} \Gamma_\mu
  (\omega) \\ 
     \times 
  \Big( Q_{j_\mu} (\omega) \varrho (t) Q_{k_\mu}^\dagger
  (\omega') - Q_{k_\mu}^\dagger (\omega')  Q_{j_\mu} (\omega) 
  \varrho (t) \Big) 
    + \mathrm{h.c.}  \,  ,
\end{multline}
where $\Gamma_\mu (\omega)$ is the one-sided Fourier transform  
\begin{equation}
  \label{eq:kerneltransform}
  \Gamma_\mu (\omega) 
  \equiv \int_0 ^\infty \mathrm{d} \tau e^{i \omega 
    \tau}  K_\mu (\tau)
\end{equation}
of the bath correlation function $K_\mu(\tau)$ given in
Eq.~\eqref{eq:kernel}. One usually neglects the Cauchy principal 
value of the integral and can then rewrite \eqref{eq:kerneltransform}
as   
\begin{equation}
  \label{eq:kerneltransform1}
    \Gamma_\mu (\omega) = 
      \begin{cases}
        J_\mu (\omega)\,(n_\mu(\omega)+1)
        \qquad &\omega \geq 0 \\ 
        J_\mu ({|}\omega{|})\, n_\mu(\omega)
        \qquad &\omega < 0 \, , 
  \end{cases}\vspace{1ex}
\end{equation}
with the spectral density $J _\mu (\omega)$ and the Bose distribution
function $n_\mu(\omega)=  1/(e^{\hbar  \omega / k_\mathrm{B} T_\mu}-1)$,
depending on the temperatures $T_\mu$.

Inserting the explicit expressions for the spectral decompositions
into the quantum master equation \eqref{eq:blochredfield1}, we find
two different classes of oscillating terms. The first oscillate at 
high frequencies such as $e^{\pm i \tilde \Omega_{A/B} t}, e^{\pm i
\omega_\mathrm{qb} t}, e^{\pm i \Delta t}, e^{\pm i \Sigma t}$, i.\,e. 
vary on timescales of the intrinsic system evolution, whereas the second
oscillate slowly at frequencies $\lambda^2_\Delta
\Delta + \lambda^2_\Sigma \Sigma $ and multiples. This difference
enables one to perform a semi-secular approximation similar to 
the approach in Ref.~\onlinecite{Scala2007a}. Here, we assume that all 
rapidly oscillating terms of the first class can be averaged to zero.
This is justified since the timescales of intrinsic 
system evolution given by $(\tilde \Omega_{A/B}) ^{-1}$ etc. are
typically much smaller than the relaxation timescales, on which the
system state varies notably. This, however, is not the case 
for terms of the second class, which we keep consistently. 
We emphasize that our approach goes beyond the standard
way of obtaining a Lindblad quantum master equation. The latter would
imply a full secular approximation, neglecting \textit{all}
oscillating contributions and only keeping terms with
$\omega= \omega'$ in Eq.~\eqref{eq:blochredfield1}. 

Furthermore, we may simplify the bath correlation functions, 
\begin{equation*}
  \begin{split}
    & \Gamma (\tilde \Omega_A(0)) \approx \Gamma (\tilde \Omega_A(1))
    \approx \Gamma
    (\tilde \Omega_B) \approx \Omega \\
    & \Gamma \big(\omega _\mathrm{qb} + n ( \lambda^2_\Delta \Delta +
    \lambda^2_\Sigma \Sigma)\big) \approx  \Gamma (\omega
    _\mathrm{qb})
  \end{split}
\end{equation*}
for small occupation numbers $n$, and assume an overall
temperature $T= T_{\{A,B,x,z\}}$. In the low-temperature regime, 
$T\ll (\hbar/k_\mathrm{B})\min\{\omega_\mathrm{qb}, \Delta,
\Omega\}$, it is appropriate to neglect all contributions to 
Eq.~\eqref{eq:blochredfield1} with negative frequencies because of 
$\Gamma_\mu (\omega< 0)\approx 0$, i.\,e. no energy is absorbed from 
the baths. This automatically yields 
$\Gamma_\mu (\omega)= J_\mu (\omega)$. In the low-frequency region
of the qubit baths, we assume Ohmic spectral behavior, $J_{\{x,z\}}
(\omega) = \alpha_{\{x,z\}} \omega$. As detailed in
App.~\ref{sec:qubitqme}, this implies $\Gamma_{\{x,z\}}
(\omega\rightarrow 0)= \alpha_{\{x,z\}} k_\mathrm{B} 
T  /\hbar$.   

We eventually obtain the effective quantum master equation for the
reduced system state  
\begin{widetext}
  \begin{equation}
    \label{eq:effQME}
    \begin{split}
      \dot \rho = &-\frac{i}{\hbar} \big[\mathcal H
      _\mathrm{eff}, \rho \big] + J_A (\Omega) D [a] \rho  + J_B
      (\Omega) D [b] \rho + (\lambda_\Delta + \lambda_\Sigma)^2 (J_x
      (\Omega) \cos^2 \theta + J_z (\Omega) \sin^2 \theta)
      D\big[\sigma_z (a+b) \big] \rho 
      \\
      &+ J_x (\omega _\mathrm{qb}) D\Big[\sigma^- \Big(\cos \theta
      - \Big(\cos \theta ( - 4 \lambda^2_\Omega + (\lambda_\Delta 
      + \lambda_\Sigma)^2) + 4 \sin \theta
      \lambda_\Omega(\lambda_\Delta
      -  \lambda_\Sigma)\Big) \Big((a^\dagger +
      b^\dagger)(a+b)+1 \Big)\Big)  \Big] \rho \\
      &+  J_x (\omega _\mathrm{qb})  D\Big[\sigma^-
      \Big(-\sin \theta + \Big(\sin \theta ( - 4 \lambda_\Omega^2 +
      (\lambda_\Delta + \lambda_\Sigma)^2) - 4 \cos \theta 
      \lambda_\Omega (\lambda_\Delta - \lambda_\Sigma) \Big) \Big((a^\dagger +
      b^\dagger)(a+b)+1 \Big)\Big)  \Big] \rho \\
      &+ 4 \big( J_x (\Delta)  (\lambda_\Omega \cos \theta +
      \lambda_\Delta \sin \theta)^2 + J_z (\Delta) 
      (\lambda_\Delta \cos \theta - \lambda_\Omega \sin \theta)^2\big)
      D\Big[\sigma^- (a^\dagger + b^\dagger)\Big] \rho \\
      &+ 4 \big(  J_x (\Sigma)   (\lambda_\Omega \cos \theta -
      \lambda_\Sigma \sin \theta)^2 +  J_z (\Sigma) 
      ( \lambda_\Sigma \cos \theta + \lambda_\Omega \sin \theta)^2 \big)
      D\Big[\sigma^- (a + b)\Big] \rho \\ 
      &+ (k_\mathrm{B} T/\hbar) (\alpha_x \sin^2 \theta + \alpha_z
      \cos^2   \theta) D\Big[\sigma_z \Big(1-2(\lambda_\Delta^2 + 
      \lambda_\Sigma ^2)
      \Big( (a^\dagger + b^\dagger)(a+b)+1\Big)\Big) \Big] \rho \; ,
     \end{split}
  \end{equation}
\end{widetext}
where we have omitted the time dependence of the
density operator $\rho$ and used the notation 
$D[X]\rho= X \varrho X^\dagger\!-\!\frac{1}{2}\big[X^\dagger X, 
\varrho \big]_+ $ with the anti-commutator $[\mathcal A,\mathcal 
B]_+= \mathcal A \mathcal B +  \mathcal B \mathcal A$.  

We have neglected terms $\propto \lambda_{\{\Delta, \Sigma,
\Omega\}}^2  J_{\{A,B\}} (\omega)$ 
and higher orders of $\lambda_{\{\Delta, \Sigma,\Omega\}}$ in
\eqref{eq:effQME}. This is justified, since the resonators typically
employed in experiment possess a high quality factor and therefore
feature small decay rates. 
On the other hand, it is necessary to keep terms proportional to 
$\lambda_{\{\Delta, \Sigma, \Omega\}}^2 J_{\{x,z\}} (\omega)$, because
typical qubit dephasing and relaxation rates exceed those of the
resonators by several orders of magnitude with conservative estimates.   
In general, it is appropriate to neglect all terms of the order of 
$\lambda_{\{\Delta, \Sigma, \Omega\}}^3$ and higher. Due to the
dissipator relation $D[\lambda X]= \lambda^2 D[X]$ we thus may
already discard any contributions to the system-bath coupling
operators of order $\lambda_{\{\Delta, \Sigma, \Omega\}}^2$ in 
Eqs.~\eqref{eq:XAeff}-\eqref{eq:sigmazeff}.  

In Sec.~\ref{sec:quantumswitch} we have motivated the experimental
advantage of keeping the qubit in its ground state without own
dynamics. This enables us to simplify \eqref{eq:effQME} further by
tracing out the qubit degrees of freedom. To this end, we take into
account  $\mathrm{Tr} \big( \sigma_z {|}g\rangle \langle g {|} 
\big) = {-} 1$ and the partial trace relation 
\begin{equation}
\rm{Tr}_1 \big(D [A_1 \otimes A_2]\rho\big) =
 \rm{Tr}_1 \big(A_1^\dagger A _1 ^{\vphantom{\dagger}} 
 D[A_2] \rho\big) \, ,
\label{eq:trace}
\end{equation}
with $A_1$ and $A_2$ acting each on a different Hilbert subspace, and
$\rm{Tr}_1 (\cdot)$ denoting the partial trace with respect to one 
subspace. Finally, 
we arrive at the effective quantum master equation for the
tow-resonator state, Eq.~\eqref{eq:effQMEcav}. The qubit decoherence
rates $\gamma_x$ and $\gamma_\phi$ are identified following the
discussion in App.~\ref{sec:qubitqme}.

\end{document}